\documentclass[preprintnumbers]{revtex4}
\usepackage{eurosym}
\usepackage{amsfonts}
\usepackage{amsmath}
\usepackage{hyperref}
\usepackage{amssymb}
\usepackage{subfigure}
\usepackage[english]{babel}
\usepackage{graphicx}
\usepackage{subfigure}
\usepackage{epsfig}
\usepackage{bm}
\usepackage{longtable}
\usepackage{verbatim}
\usepackage{longtable}
\usepackage[utf8]{inputenc}
\usepackage{xcolor}
\usepackage{wrapfig}
\pdfoutput=1

\newcommand{\mathsym}[1]{{}}

\input{tcilatex}
\begin{document}

\title{Sequentially loop-generated quark and lepton mass hierarchies in an
extended Inert Higgs Doublet model}
\author{A. E. C\'arcamo Hern\'andez$^{{a}}$}
\email{antonio.carcamo@usm.cl}
\author{Sergey Kovalenko$^{{a}}$}
\email{sergey.kovalenko@usm.cl}
\author{Roman Pasechnik{}$^{b,c,d}$}
\email{Roman.Pasechnik@thep.lu.se}
\author{Ivan Schmidt$^{{a}}$}
\email{ivan.schmidt@usm.cl}
\affiliation{$^{{a}}$Universidad T\'ecnica Federico Santa Mar\'{\i}a and Centro Cient%
\'{\i}fico-Tecnol\'ogico de Valpara\'{\i}so\\
Casilla 110-V, Valpara\'{\i}so, Chile\\
$^{{b}}$Department of Astronomy and Theoretical Physics, \\
Lund University, Solvegatan 14A, SE-223 62 Lund, Sweden\\
$^{{c}}$Nuclear Physics Institute ASCR, 25068 \v{R}e\v{z}, Czech Republic\\
$^{{d}}$Departamento de F\'isica, CFM, Universidade Federal\\
de Santa Catarina, C.P. 476, CEP 88.040-900, Florian\'opolis, SC, Brazil }

\begin{abstract}
Extended scalar and fermion sectors offer new opportunities for generating
the observed strong hierarchies in the fermion mass and mixing patterns of
the Standard Model (SM). In this work, we elaborate on the prospects of a
particular extension of the Inert Higgs doublet model where the SM
hierarchies are generated sequentially by radiative virtual corrections in a
fully renormalisable way, i.e. without adding any non-renormalisable Yukawa
terms or soft-breaking operators to the scalar potential. Our model has a
potential to explain the recently observed $R_{K}$ and $R_{K^{\ast }}$
anomalies, thanks to the non universal $U_{1X}$ assignments of the fermionic
fields that yield non universal $Z^{\prime}$ couplings to fermions. We
explicitly demonstrate the power of this model for generating the realistic
quark, lepton and neutrino mass spectra. In particular, we show that due to
the presence of both continuous and discrete family symmetries in the
considered framework, the top quark acquires a tree-level mass, lighter
quarks and leptons get their masses at one- and two-loop order, while
neutrino masses are generated at three-loop level. The minimal field
content, particle spectra and scalar potential of this model are discussed
in detail.
\end{abstract}

\maketitle



\section{Introduction}

\label{Sec:Intro} 

The origin of various strong hierarchies in the fermion spectra of the
Standard Model (SM) still remains a major unsolved problem of contemporary
Particle Physics. A symmetry-based understanding of such hierarchies, in the
framework of a single universal mechanism, consistent with the current
phenomenological bounds on New Physics models, poses a challenging problem
for the model-building community. In fact, a number of different potentially
realizable mechanisms have been proposed so far, typically with certain
limitations and deficiencies, and usually treating the quark, lepton and
neutrino hierarchies on a separate footing. The most promising scenarios
rely on the existence of horizontal (family) symmetries acting in the space
of fermion generations and offering most times a more universal approach to
the ``fermion hierarchy problem'' than other methods.

It seems natural and attractive to consider the viable prospect of a
high-scale spontaneous discrete symmetry breaking, triggering the radiative
generation of new mass operators in the corresponding low-energy effective
field theory (EFT) \cite
{Balakrishna:1988ks,Ma:1988fp,Ma:1989ys,Kitabayashi:2000nq,Chang:2006aa,Hernandez:2013mcf,Hernandez:2013dta,Campos:2014lla,Boucenna:2014ela,Okada:2015bxa,Wang:2015saa,Arbelaez:2016mhg,Nomura:2016emz,Kownacki:2016hpm,Nomura:2016ezz,Hernandez:2015hrt,Camargo-Molina:2016yqm,Camargo-Molina:2016bwm,CarcamoHernandez:2017kra,Abbas:2017vws,Dev:2018pjn,CarcamoHernandez:2018hst,Abbas:2018lga}%
. This way, one arrives at the possible sequential generation (in general,
due to a few sequential symmetry breakings at the high-energy scales) of the
relevant mass (or SM Higgs Yukawa) terms in the SM, whose values are matched
to zero or to a universal non-zero value in the high-scale limit of a more
symmetric, and hence more fundamental theory. The search for such
ultraviolet (UV) completions, possessing a low-energy SM-like EFT, and
explorations of their vast potential for explaining the origin of the SM
structure have only began recently \cite%
{CarcamoHernandez:2016pdu,CarcamoHernandez:2017cwi}.

A model of radiatively generated fermion masses in the SM via sequential
loop suppression has been proposed in Ref.~\cite{CarcamoHernandez:2016pdu}.
In this model, the top quark mass is generated at tree level, while the
bottom, tau and muon lepton masses arise at one-loop level. Meanwhile, the
smaller up, down and strange quark and electron masses are generated at
two-loop level, while the light active neutrinos acquire their masses at
four-loop level. However, such a natural sequential generation of fermion
masses at various orders of the Perturbation Theory comes with a price.
Namely, this model, based on the SM gauge symmetry, supplemented with the $%
S_{3}\times Z_{2}$ discrete group, has a quite low cut-off, since it
incorporates non-renormalizable Yukawa terms. In addition, it has another
drawback in that the $S_{3}\times Z_{2}$ discrete group is softly broken,
yielding an unknown UV completion of the theory. This situation may indicate
the need for horizontal continuous symmetries for a consistent description
of hierarchies in the matter sectors in a fully renormalizable way.

As a follow up to this study, in the later work of Ref.~\cite%
{CarcamoHernandez:2017cwi}, we proposed such a renormalizable model based on
the $SU_{3C}\times SU_{3L}\times U_{1X} \times U_{1\mathcal{L}}\times
Z_{2}^{\left( 1\right) }\times Z_{2}^{\left( 2\right) }$ symmetry, which generates the SM fermion mass hierarchies
with an emergent sequential loop suppression mechanism. However, that model
still retains some drawbacks of the previous formulation. Namely, the $%
Z_{2}^{\left( 1\right) }\times Z_{2}^{\left( 2\right) }$ symmetry is softly broken, and the scalar and fermion
sectors are excessively large, making it very difficult to perform a
reliable phenomenological analysis. In addition, that model does not explain
the $R_{K}$ and $R_{K^{\ast }}$ anomalies, recently observed by the LHCb
experiment \cite{Aaij:2014ora,Aaij:2013qta,Aaij:2015oid}, since it treats
the first and second lepton families in the same footing. Furthermore, in
the model of Ref.~\cite{CarcamoHernandez:2017cwi}, the masses for the light
active neutrinos appear at two-loop level, as well as the masses for the
electron and for the light up, down and strange quarks. Thus the smallness
of the light active neutrino masses with respect to the electron mass does
not receive a natural explanation in the model. As a natural step forward,
it would be instructive to find a new analogous formulation that enables us
to generate the SM charged fermion mass hierarchies via a sequential loop
suppression mechanism and to generate three-loop level light active neutrino
masses without the inclusion of soft breaking mass terms. It would also be
relevant to further explore the potential of such formulations for
explaining the LHCb anomalies.

In the current work, we propose a first renormalizable model, an extended
variant of the Inert Higgs Doublet model (IDM) \cite{Deshpande:1977rw}, that
enables to generate strong fermion hierarchies via another sequential loop
suppression pattern, not yet discussed in the literature, without
introducing any soft family symmetry breaking mass terms. Similarly to the
previous formulations, in the current model the top quark and exotic
fermions do acquire tree level masses, whereas the masses of the remaining
SM fermions are radiatively generated. Namely, the masses for the bottom,
strange and charm quarks, tau and muon leptons are generated at one-loop
level, whereas the masses for the up and down quarks, as well as the
electron mass, arise at two-loop level. In variance to the previous version
in Ref.~\cite{CarcamoHernandez:2016pdu}, the light active neutrinos acquire
masses via radiative seesaw mechanisms at three-loop level, whereas in the
former the light active neutrino masses were induced at two-loop level.
Finally, the minimal field content of the model is not as complicated as in
the previous formulations, enabling us to explore several key
phenomenological implications of this model, which is the main subject of a
follow up study.


The current article is organized as follows. In section~\ref{Sec:sequential}
we discuss generic conditions for a sequential loop suppression mechanism,
providing a motivation for the proposed model. In section~\ref%
{Sec:Extended-IDM}, we set up the formalism for the extended IDM, containing
the basic details about the symmetries, particle content, Yukawa
interactions and scalar potential crucial for the implementation of the
sequential loop suppression mechanism. The scalar mass spectrum is discussed
in detail in section~\ref{Sec:scalar-mass}. In section~\ref%
{Sec:quark-mass-mixing} we discuss the implications of our model for the
radiative generation of the quark mass and mixing hierarchies. The charged
lepton and neutrino mass spectra and the corresponding mixing patterns are
discussed in section~\ref{Sec:lepton-mass-mixing}. Finally, our conclusions
are briefly stated in section~\ref{Sec:conclusions}.


\section{Sequential loop suppression mechanism}

\label{Sec:sequential} 

Before describing our model in detail, let us first explain the reasoning
behind introducing the additional scalar and fermion degrees of freedom and
the symmetries that are required for a consistent implementation of the
sequential loop suppression mechanism for generating the SM fermion
hierarchies.


\subsection{Quark sector}

\label{Sec:quarks} 

First of all, it is worth noticing that the top quark mass can be generated
at tree level by means of a renormalizable Yukawa operator, with the
corresponding coupling of order one, i.e. 
\begin{equation}
\overline{q}_{3L}\widetilde{\phi }_{1}u_{jR},\hspace{1cm}j=1,2,3,
\end{equation}%
where $\phi _{1}$ is a $SU_{2L}$ scalar doublet. To generate the charm quark
mass at one loop level, it is necessary to forbid the operator: 
\begin{equation}
\overline{q}_{nL}\widetilde{\phi }_{1}u_{jR},\hspace{1cm}n=1,2,\hspace{1cm}%
j=1,2,3,
\end{equation}%
at tree level and to allow other operators instead, which are described in
what follows. Obviously, in this case the charm quark mass is not generated
in the same way as the top quark mass, i.e., from a renormalizable Yukawa
term, since this would imply setting the corresponding tree-level Yukawa
coupling unnaturally small.

In order to generate a small charm quark mass at one-loop level, we will use
the following operators: 
\begin{eqnarray}
&&\overline{q}_{nL}\widetilde{\phi }_{2}T_{R},\hspace{1cm}\hspace{1cm}%
\overline{T}_{L}\eta ^{\ast }u_{mR},\hspace{1cm}\hspace{1cm}n,m=1,2,  \notag
\\
&&\overline{T}_{L}\sigma _{1}T_{R},\hspace{1cm}\hspace{1cm}\sigma
_{2}^{\ast }\rho _{2}^{\ast }\sigma _{1}\rho _{3}^{\ast },\hspace{1cm}%
\hspace{1cm}\eta \rho _{2}\sigma _{1},\hspace{1cm}\hspace{1cm}\left( \phi
_{1}^{\dagger }\cdot \phi _{2}\right) \sigma _{2}\,.
\end{eqnarray}%
For this to happen, we need to extend the SM gauge symmetry by adding the $%
U_{1X}\times Z_{2}^{\left( 1\right) }\times Z_{2}^{\left( 2\right) }$
symmetry, where the $U_{1X}$ and $Z_{2}^{\left( 1\right) }$ are
spontaneously broken gauge and discrete symmetries, respectively, and $%
Z_{2}^{\left( 2\right) }$ is a preserved (exact) discrete symmetry under
which the extra $\phi _{2}$ scalar doublet, the $\sigma _{2}$, $\rho _{2}$, $%
\eta $ electrically neutral scalar singlets, and the $q_{jL}$, $u_{jR}$ ($%
j=1,2,3$) quark fields are nontrivially charged. It is worth mentioning that
the fields $u_{nR}$ ($n=1,2$), $\rho _{2}$, $\rho _{3}$\ and $\eta $ are
charged under the spontaneously broken $Z_{2}^{\left( 1\right) }$ symmetry,
whereas the remaining fields previously introduced are neutral under this
symmetry. As we will be shown below, the $\sigma _{2}$, $\rho _{2}$, $\eta $
electrically neutral scalar singlets will also mediate the one and two level
radiative seesaw mechanisms that give masses to the second and third family
of SM down type quarks and charged leptons as well as to the first family of
SM charged fermions. Some of these scalars singlets will also participate in
the three loop level radiative seesaw mechanism that produces the light
active neutrino masses. Furthermore, the $SU_{2L}$ singlet heavy quarks $%
T_{L}$, $T_{R}$ with electric charges equal to $2/3$ have to be added to the
fermion spectrum in order to implement the one-loop radiative seesaw
mechanism that gives rise to the charm quark mass. Besides, an electrically
neutral weak-singlet scalar $\sigma _{1}$ is needed in the scalar spectrum
to provide a tree-level mass for the exotic $T$ quark and to close the
one-loop Feynman diagram that generates the charm quark mass. Let us note
that $\sigma _{1}$ and $\rho _{3}$ are the only scalars neutral under the
unbroken $Z_{2}^{\left( 2\right) }$ symmetry, and thus they conveniently
acquire vacuum expectation values (VEVs) that break the $U_{1X}$ gauge
symmetry and the $Z_{2}^{\left( 1\right) }$ discrete symmetry, respectively.

In addition, we assume that the $q_{nL}$ ($n=1,2$) fields have $U_{1X}$
charges that are different from the charge of $q_{3L}$, while the $SU_{2L}$
scalar doublets $\phi _{1}$ and $\phi _{2}$ have different $U_{1X}$ charges
as well. Thus, the third row of the up-type quark mass matrix is generated
at tree level, whereas the first and second row emerge at one-loop level.
Note that, since there is only one heavy exotic $T$ quark mediating the
one-loop radiative seesaw mechanism that generates the first and second row
of the up-type quark mass matrix, the determinant of this matrix is equal to
zero. Therefore, the up quark is massless at one-loop level, and in order to
generate an up quark mass at two-loop level, the following operators are
required: 
\begin{eqnarray}
&&\overline{q}_{nL}\phi _{2}B_{3R},\hspace{1cm}\hspace{1cm}\overline{B}%
_{4L}\varphi _{1}^{-}u_{nR},\hspace{1cm}\hspace{1cm}\overline{B}_{3L}\sigma
_{2}^{\ast }B_{4R},\hspace{1cm}\hspace{1cm}m_{B_{k}}\overline{B}_{kL}B_{kR},%
\hspace{1cm}\hspace{1cm}n=1,2,\hspace{1cm}  \notag \\
&&\sigma _{1}\sigma _{2}\sigma _{3}^{\ast }\rho _{3},\hspace{1cm}\hspace{1cm}%
\varepsilon _{ab}\left( \phi _{1}^{\dagger }\right) ^{a}\left( \phi
_{2}^{\dagger }\right) ^{b}\varphi _{1}^{+}\sigma _{3},\hspace{1cm}\hspace{%
1cm}a,b=1,2\,,,\hspace{1cm}\hspace{1cm}\,k=3,4,
\end{eqnarray}%
where the extra $SU_{2L}$ singlet heavy quarks $B_{3L}$, $B_{3R}$, $B_{4L}$, 
$B_{4R}$ have electric charges equal to $-1/3$, and $\sigma _{3}$ and $%
\varphi _{1}^{-}$ are electrically neutral and charged weak-singlet scalars,
respectively. The scalar fields $\sigma _{3}$ and $\varphi _{1}^{-}$ are
charged under the spontaneously broken $Z_{2}^{\left( 1\right) }$ symmetry.
The operators given above enable to generate the two-loop contributions to
the first and second rows of the up-type quark mass matrix, and these
contributions yield a nonvanishing determinant for the up-type quark mass
matrix, giving rise to a suppressed two-loop up quark mass. It is worth
mentioning that the nonvanishing parts of the SM up type quark mass matrix
top quark are the $\left( 3,3\right) $ entry, which appears at tree level
and the upper left $2\times 2$ block which receive one and two loop level
contributions. 
Let us note that in order that the two loop contribution to
the upper left $2\times 2$ block of the SM up type quark mass matrix gives a
nonvanishing two loop level up quark mass, the one and two loop level contributions should not have common vertices. If they have a common vertex, the two loop level contribution can be
treated as an effective 1-loop one having the same the left and (or) right
handed fermionic field in the internal line as in the former one loop level
contribution. Thus, the net result of the sum of both contributions will
correspond to one loop level diagram having in the fermionic line a seesaw
mediator, which will be linear combination of the fermionic mediators in
both contributions, thus leading to a vanishing up quark mass. Another
reason for avoiding a common vertex in the one and two loop level
contributions to the upper left $2\times 2$ block of the SM up type quark
mass matrix, is to prevent a proportionality between row and columns that
will result in a vanishing eigenvalue. Because of the aforementioned reason,
we have employed exotic down type quarks and electrically charged scalars in
the two loop level contribution of the upper left $2\times 2$ block of the
SM up type quark mass matrix, thus giving rise to a up quark mass at two
loop level. As it will be shown below, the two loop level contributions to
the SM charged fermion mass matrices, will be generated by electrically
charged scalars and exotic fermions running in the internal lines of the
loop, thus yielding two loop level masses for the first generation of SM
charged fermions.  

Turning now to a possible bottom quark mass generation at one-loop level,
the following operators should be forbidden 
\begin{equation}
\overline{q}_{3L}\phi _{1}d_{jR}\,,\hspace{1cm}\hspace{1cm}j=1,2,3\,,
\end{equation}%
by means of, for example, the $U_{1X}$ gauge symmetry. The fermion spectrum
has to be extended by including the additional $SU_{2L}$-singlet heavy
quarks $B_{nL}$, $B_{nR}$ ($n=1,2$) with electric charges equal to $-1/3$,
so that the mass for the bottom quark is generated with the help of the
following operators: 
\begin{eqnarray}
&&\overline{q}_{3L}\phi _{2}B_{nR},\hspace{1cm}\hspace{1cm}\overline{B}%
_{nL}\eta d_{jR},\hspace{1cm}\hspace{1cm}n=1,2,\hspace{1cm}\hspace{1cm}%
n,m=1,2,\hspace{1cm}\hspace{1cm}j=1,2,3,  \notag \\
&&\overline{B}_{nL}\sigma _{1}^{\ast }B_{mR},\hspace{1cm}\hspace{1cm}\sigma
_{2}\rho _{2}\left( \sigma _{1}^{\ast }\right) \rho _{3},\hspace{1cm}\hspace{%
1cm}\eta ^{\ast }\rho _{2}^{\ast }\sigma _{1}^{\ast },\hspace{1cm}\hspace{1cm%
}\left( \phi _{1}\cdot \phi _{2}^{\dagger }\right) \sigma _{2}^{\ast }\,.
\end{eqnarray}%
Note that we have added two $SU_{2L}$-singlet heavy quarks $B_{n}$ ($n=1,2$)
instead of one, in order to fulfill the anomaly cancellation conditions
discussed below.

In addition, in order to generate the first and second rows of the down-type
quark mass matrix at one-loop level via a radiative seesaw mechanism, we
need the following set of operators: 
\begin{eqnarray}
&&\overline{q}_{nL}\phi _{2}B_{3R},\hspace{1cm}\hspace{1cm}\overline{B}%
_{3L}\eta ^{\ast }d_{jR},\hspace{1cm}\hspace{1cm}n=1,2,\hspace{1cm}\hspace{%
1cm}j=1,2,3,  \notag \\
&&m_{B_{3}}\overline{B}_{3L}B_{3R},\hspace{1cm}\hspace{1cm}\eta \left( \phi
_{1}\cdot \phi _{2}^{\dagger }\right) \rho _{3},\hspace{1cm}\hspace{1cm}
\end{eqnarray}

Furthermore, in order to avoid tree-level masses for the down and strange
quarks one has to forbid the terms 
\begin{equation}
\overline{q}_{nL}\phi _{1}d_{jR},\hspace{1cm}n=1,2,\hspace{1cm}j=1,2,3\,.
\end{equation}%
The latter can be excluded by assigning $q_{nL}$ ($n=1,2$) to be even, while 
$d_{jR}$ ($j=1,2,3$) to be odd under the aforementioned spontaneously broken 
$Z_{2}^{\left( 1\right) }$symmetry.

Since there is only one fermionic seesaw mediator, i.e., the $SU_{2L}$
singlet heavy quark $T$ needed to generate the first and second rows of the
down-type quark mass matrix at one-loop level, a nonvanishing one-loop
strange quark mass emerges, whereas the down quark remains massless at this
point. Consequently, similarly to the up-type quark sector, the two-loop
contributions to the first and second rows of the down-type quark mass
matrix need to be generated in order to give rise to a down-type quark mass
at two-loop level. For that purpose, the following operators are required: 
\begin{eqnarray}
&&\overline{q}_{nL}\widetilde{\phi }_{2}T_{R},\hspace{1cm}\hspace{1cm}%
\overline{\widetilde{T}}_{L}\varphi _{2}^{+}d_{jR},\hspace{1cm}\hspace{1cm}%
m_{\widetilde{T}}\overline{\widetilde{T}}_{L}\widetilde{T}_{R},\hspace{1cm}%
\hspace{1cm}n=1,2,\hspace{1cm}\hspace{1cm}j=1,2,3, \\
&&\overline{T}_{L}\sigma _{1}T_{R},\hspace{1cm}\hspace{1cm}\overline{T}%
_{L}\rho _{2}\widetilde{T}_{R},\hspace{1cm}\hspace{1cm}\sigma _{1}\sigma
_{2}^{\ast }\rho _{2}\rho _{3},\hspace{1cm}\hspace{1cm}\varepsilon
_{ab}\left( \phi _{1}\right) ^{a}\left( \phi _{2}\right) ^{b}\varphi
_{2}^{-}\sigma _{2},\hspace{1cm}\hspace{1cm}a,b=1,2\,,  \notag
\end{eqnarray}%
where an extra electrically charged weak-singlet scalar, $\varphi _{2}^{+}$
has been added to the scalar spectrum. Furthermore, the fermion spectrum has
been extended by means of extra $SU_{2L}$-singlet heavy quarks $\widetilde{T}%
_{L}$, $\widetilde{T}_{R}$ with electric charges equal to $2/3$. The
two-loop contributions to the first and second rows of the down-type quark
mass matrix provide its nonvanishing determinant, giving rise to a two-loop
down quark mass.


\subsection{Charged lepton sector}

\label{Sec:leptons} 

Now, consider the sequential loop suppression mechanism capable of
explaining the observed hierarchy between the SM charged lepton masses. In
what follows, let us discuss a possible way of generating the one-loop tau
and muon masses as well as a two-loop electron mass. First of all, the
following operators have to be forbidden 
\begin{equation}
\overline{l}_{iL}\phi _{1}l_{jR},\hspace{1cm}\hspace{1cm}i,j=1,2,3.  \notag
\end{equation}

by using the $U_{1X}$ gauge symmetry, so that the SM charged lepton mass
matrix is generated at one-loop level by means of the terms 
\begin{eqnarray}
&&\overline{l}_{kL}\phi _{2}E_{3R},\hspace{1cm}\hspace{1cm}\overline{E}%
_{3L}\rho _{1}l_{nR},\hspace{1cm}\hspace{1cm}\left( \phi _{1}\cdot \phi
_{2}^{\dagger }\right) \sigma _{2}^{\ast },\hspace{1cm}\hspace{1cm}\rho
_{1}\sigma _{2}\sigma _{1}^{\ast },\hspace{1cm}\hspace{1cm}n=1,3,  \notag
\\
&&\overline{l}_{2L}\phi _{2}E_{2R},\hspace{1cm}\hspace{1cm}\overline{E}%
_{2L}\rho _{1}l_{2R},\hspace{1cm}\hspace{1cm}\overline{E}_{jL}\sigma
_{1}^{\ast }E_{jR},\hspace{1cm}\hspace{1cm}j=2,3\,,
\end{eqnarray}%
where weak-singlet charged leptons $E_{jL}$, $E_{jR}$ ($j=2,3$) have been
included in the fermion spectrum. Let us note that these fields mediate the
one-loop radiative seesaw mechanism that generates the $\left( 1,1\right) $, 
$\left( 2,2\right) $, $\left( 3,3\right) $, $\left( 1,3\right) $ and $\left(
3,1\right) $ entries of the charged lepton mass matrix. Consequently, at
this point the determinant of the charged lepton mass matrix is equal to
zero and thus only the tau and muon leptons appear to be massive at one-loop
level, whereas the electron remains massless. Two-loop corrections to the $%
\left( 1,1\right) $, $\left( 3,3\right) $, $\left( 1,3\right) $ and $\left(
3,1\right) $ entries are needed in order to induce a non-zero electron mass
at two-loop level, and extra weak-singlet charged $E_{1}$ and neutral $\nu
_{mR}$ ($m=1,3$), \ leptons $\Psi _{R}$ as well as the electrically charged
scalar singlets $\varphi _{1}^{\pm }$, $\varphi _{k}^{\pm }$\ \ ($k=3,4,5$)
would be required for this purpose. Let us note that the neutral gauge
singlet leptons $\nu _{mR}$ ($m=1,3$), \ leptons $\Psi _{R}$ will also
participate in the three loop leval radiative seesaw mechanism that produces
the small light active neutrino masses. To obtain the two-loop corrections
of the SM charged lepton mass matrix that generate a two loop level electron
mass, the following operators should be considered 
\begin{eqnarray}
&&\overline{l}_{kL}\widetilde{\phi }_{2}\nu _{nR},\hspace{1cm}\hspace{1cm}%
\overline{\Psi _{R}^{C}}\varphi _{3}^{+}l_{kR},\hspace{1cm}\hspace{1cm}%
\varphi _{4}^{-}\varphi _{5}^{+}\sigma _{1},\hspace{1cm}\hspace{1cm}\varphi
_{1}^{+}\varphi _{5}^{-}\sigma _{1}^{2},\hspace{1cm}\hspace{1cm}k,n=1,3, \\
&&\overline{E}_{1L}\sigma _{1}^{\ast }E_{1R},\hspace{1cm}\hspace{1cm}%
\overline{E}_{1L}\varphi _{1}^{-}\nu _{kR},\hspace{1cm}\hspace{1cm}\overline{%
\Psi _{R}^{C}}\varphi _{4}^{+}E_{1R},\hspace{1cm}\hspace{1cm}\varepsilon
_{ab}\left( \phi _{1}\right) ^{a}\left( \phi _{2}\right) ^{b}\varphi
_{3}^{-},\hspace{1cm}\hspace{1cm}a,b=1,2,  \notag
\end{eqnarray}%
As soon as the two-loop corrections generated by the above operators are
included, the determinant of the charged lepton mass matrix becomes nonzero,
thus giving rise to a two-loop electron mass, as expected.

Let us note that in this setup there is a mixing between the SM charged
leptons of the first and third generations, but they do not mix with the
second generation, which is a consequence of their $U_{1X}$ assignments, as
will be shown below. The second generation left-handed lepton $SU_{2L}$%
-doublet has a nonvanishing $U_{1X}$ charge, whereas the first and third
generations are not charged under $U_{1X}$. In addition, the first and third
generations of SM right-handed charged leptons should have the same $U_{1X}$
charge, which is different from the corresponding charge of the second
generation.


\subsection{Light active neutrino sector}

\label{Sec:neutrino} 

Turning to the neutrino sector, in order to generate the SM light active
neutrino masses at three-loop level as well as a realistic lepton mixing, a
few operators have to be forbidden, namely, 
\begin{equation}
\overline{l}_{iL}\phi _{1}\nu _{jR},\hspace{1.3cm}\left( m_{N}\right)
_{ij}\nu _{iR}\overline{\nu _{jR}^{C}},\hspace{1.3cm}\nu _{iR}\sigma _{1}%
\overline{\nu _{jR}^{C}},\hspace{1.3cm}\nu _{iR}\sigma _{1}^{\ast }\overline{%
\nu _{jR}^{C}}\hspace{1cm}m_{\Omega }\overline{\Omega _{R}^{C}}\Omega _{R},%
\hspace{1.3cm}i,j=1,2,3\,,
\end{equation}%
while the following operators are required 
\begin{eqnarray}
&&\overline{l}_{kL}\widetilde{\phi }_{2}\nu _{nR},\hspace{1cm}\hspace{1cm}%
\overline{l}_{2L}\widetilde{\phi }_{2}\nu _{2R},\hspace{1cm}\hspace{1cm}%
\overline{\Omega _{1R}^{C}}\eta ^{\ast }\nu _{kR},\hspace{1cm}\hspace{1cm}%
\overline{\Omega _{1R}^{C}}\sigma _{3}^{\ast }\nu _{2R},\hspace{1cm}\hspace{%
1cm}k,n=1,3\,.  \notag \\
&&\overline{\Omega _{1R}^{C}}\eta \Psi _{R},\hspace{1cm}\hspace{1cm}%
\overline{\Omega _{2R}^{C}}\eta ^{\ast }\Psi _{R},\hspace{1cm}\hspace{1cm}%
\overline{\Omega _{1R}^{C}}\sigma _{2}^{\ast }\Omega _{2R},\hspace{1cm}%
\hspace{1cm}m_{\Psi }\overline{\Psi _{R}^{C}}\Psi _{R}\,,
\end{eqnarray}%
where $\nu _{iR}$ ($i=1,2,3$), $\Omega _{R}$ and $\Psi _{R}$ are the
SM-singlet right-handed Majorana neutrinos, and $\eta $, $\sigma _{3}$ and $%
\rho _{2}$ are the extra SM-singlet scalars. The latter fields have to be
added in order to ensure three-loop level generation of the SM light active
neutrino masses, as well as the lepton mixing parameters $\sin ^{2}\theta
_{23}$ and $\sin ^{2}\theta _{12}$.

As was mentioned above, the charged lepton mass matrix has a mixing only in
the (1,3)-plane such that the lepton mixing parameters $\sin ^{2}\theta
_{23} $ and $\sin ^{2}\theta _{12}$ emerge from the neutrino sector. Let us
note that the $U_{1X}$ gauge symmetry prevents the light active neutrino
masses to be generated at tree level, whereas the $Z_{2}^{\left( 1\right) }$
symmetry, together with the $U_{1X}$ gauge symmetry, forbid the mixing terms
between the right-handed Majorana neutrinos $\nu _{kR}$ ($k=1,3$) and $\nu
_{2R}$. This enables us to avoid the appearance of light neutrino masses at
one- and two-loop levels. Note also that the $Z_{2}^{\left( 1\right) }$
symmetry, as well as the $U_{1X}$ gauge symmetry, are crucial to forbid the
terms $\nu _{2R}\sigma _{1}\overline{\nu _{kR}^{C}}$ and $\nu _{2R}\rho _{3}%
\overline{\nu _{kR}^{C}}$ ($k=1,3$) that could result in the appearance of
SM neutrino masses at one loop.


\section{The extended IDM model}

\label{Sec:Extended-IDM} 

In this section, we will summarize the main features of the first
renormalizable model, an extended variant of the Inert Higgs Doublet model
(IDM), that includes a sequential loop suppression mechanism for the
generation of the SM fermion mass hierarchies, without the inclusion of soft
breaking mass terms and, at the same time, allowing for an explanation of
the $R_{K}$ and $R_{K^{\ast }}$ anomalies, thanks to the non-universal $%
U_{1X}$ assignments of the fermionic fields that yield non-universal $%
Z^{\prime}$ couplings to fermions. In a forthcoming work, we will show in
detail how our model can fit the $R_{K}$ and $R_{K^{\ast }}$ anomalies. As
previously stated in Introduction, we emphasize that our model,
apart from having all the means for explaining the $R_{K}$ and $R_{K^{\ast
}} $ anomalies not previously addressed in the model of Ref.~\cite%
{CarcamoHernandez:2016pdu}, has a more natural explanation for the smallness
of the light active neutrino masses than the one provided in Ref.~\cite%
{CarcamoHernandez:2016pdu}, since in the former the masses for the light
active neutrinos are generated at three-loop level, whereas in the latter
they appear at two loops. Furthermore, unlike the model of Ref.~\cite%
{CarcamoHernandez:2016pdu}, our current model does not include soft breaking
mass terms.

Let us summarize the structure of a minimal model where the sequential loop
suppression mechanism capable of radiative generation of the mass and mixing
hierarchies in the SM fermion sectors is realized. The reasons for choosing
a particular field content and symmetries have been outlined in the previous
section, and will be further substantiated below.


\subsection{Particle content and charges}

\label{Sec:content} 

With the necessary ingredients introduced above, in fact, we arrive at an
extension of the inert 2HDM where the SM gauge symmetry is supplemented by
the exact unbroken $Z_{2}^{\left( 2\right) }$ discrete group and
spontaneously broken $Z_{2}^{\left( 1\right) }$ discrete and $U_{1X}$ gauge
symmetry groups. The unbroken $Z_{2}^{\left( 2\right) }$ and continuous
local $U_{1X}$ (horizontal) family symmetries are crucial for the
implementation of radiative seesaw mechanism of sequential loop suppression.

Besides the SM-like Higgs doublet $\phi _{1}$, the implementation of this
mechanism requires an additional inert scalar $SU_{2L}$-doublet, $\phi _{2}$%
\ , seven electrically neutral, i.e., $\sigma _{j}$, $\rho _{j}$ ($j=1,2,3$%
), $\eta $, and five electrically charged $\varphi _{k}^{+}$ ($k=1,2,3,4,5$%
), $SU_{2L}$-singlet scalars. The scalar sector of the considering model has
the following $SU_{3c}\times SU_{2L}\times U_{1Y}\times U_{1X}$ charge
assignments 
\begin{eqnarray}
\phi _{1} &\sim &\left( \mathbf{1,2},\frac{1}{2},1\right) ,\hspace{1cm}\phi
_{2}\sim \left( \mathbf{1,2},\frac{1}{2},2\right) ,\hspace{1cm}\phi _{3}\sim
\left( \mathbf{1,2},\frac{1}{2},2\right) ,\hspace{1cm}\sigma _{1}\sim \left( 
\mathbf{1,1},0,-1\right) ,\hspace{1cm}  \notag \\
\sigma _{2} &\sim &\left( \mathbf{1,1},0,-1\right) ,\hspace{1cm}\sigma
_{3}\sim \left( \mathbf{1,1},0,-2\right) ,\hspace{1cm}\rho _{1}\sim \left( 
\mathbf{1,1},0,0\right) ,\hspace{1cm}\rho _{2}\sim \left( \mathbf{1,1}%
,0,0\right) ,\hspace{1cm}  \notag \\
\rho _{3} &\sim &\left( \mathbf{1,1},0,1\right) ,\hspace{1cm}\eta \sim
\left( \mathbf{1,1},0,1\right) ,\hspace{1cm}\varphi _{1}^{+}\sim \left( 
\mathbf{1,1},1,5\right) ,\hspace{1cm}\varphi _{2}^{+}\sim \left( \mathbf{1,1}%
,1,2\right) ,\hspace{1cm}  \notag \\
\varphi _{3}^{+} &\sim &\left( \mathbf{1,1},1,3\right) ,\hspace{1cm}\varphi
_{4}^{+}\sim \left( \mathbf{1,1},1,2\right) ,\hspace{1cm}\varphi
_{5}^{+}\sim \left( \mathbf{1,1},1,3\right) \,.
\end{eqnarray}

As it will be shown in the next subsection, we provide a detailed analysis
of the scalar potential showing that the only massless scalar fields are
those ones corresponding to the SM Goldstone bosons and the Goldstone boson
associated to the longitudinal component of the $Z^{\prime }$ gauge boson.

The corresponding $Z_{2}^{\left( 1\right) }\times Z_{2}^{\left( 2\right) }$
charges of the scalar fields are given by 
\begin{eqnarray}
\phi _{1} &\sim &\left( 1,1\right) ,\hspace{1cm}\phi _{2}\sim \left(
1,-1\right) ,\hspace{1cm}\sigma _{1}\sim \left( 1,1\right) ,\hspace{1cm}%
\sigma _{2}\sim \left( 1,-1\right) ,\hspace{1cm}\sigma _{3}\sim \left(
-1,-1\right) ,  \notag \\
\rho _{1} &\sim &\left( 1,-1\right) ,\hspace{1cm}\rho _{2}\sim \left(
-1,-1\right) ,\hspace{1cm}\rho _{3}\sim \left( -1,1\right) ,\hspace{1cm}\eta
\sim \left( -1,-1\right) ,\hspace{1cm}  \notag \\
\varphi _{1}^{+} &\sim &\left( -1,1\right) ,\hspace{0.7cm}\hspace{0.7cm}%
\varphi _{2}^{+}\sim \left( 1,1\right) ,\hspace{0.7cm}\hspace{0.7cm}\varphi
_{3}^{+}\sim \left( 1,-1\right) ,\hspace{0.7cm}\hspace{0.7cm}\varphi
_{4}^{+}\sim \left( -1,1\right) \,,\hspace{0.7cm}\hspace{0.7cm}\varphi
_{5}^{+}\sim \left( -1,1\right) \,,
\end{eqnarray}%
It is worth noticing that the model does not contain a weak-singlet scalar
field with the following simultaneous three features: charged under the $%
Z_{2}^{\left( 1\right) }$ discrete symmetry, neutral under the unbroken $%
Z_{2}^{\left( 2\right) }$ symmetry with $U_{1X}$ charge equal to $\pm 1$. 
\begin{table}[th]
\centering%
\begin{tabular}{|c||c|c|c|c|c|c|c|c|c|c|c|c|c|c|}
\hline\hline
Field & $\phi _{1}$ & $\phi _{2}$ & $\sigma _{1}$ & $\sigma _{2}$ & $\sigma
_{3}$ & $\rho _{1}$ & $\rho _{2}$ & $\rho _{3}$ & $\eta $ & $\varphi
_{1}^{+} $ & $\varphi _{2}^{+}$ & $\varphi _{3}^{+}$ & $\varphi _{4}^{+}$ & $%
\varphi _{5}^{+}$ \\ \hline
$SU_{3c}$ & $\mathbf{1}$ & $\mathbf{1}$ & $\mathbf{1}$ & $\mathbf{1}$ & $%
\mathbf{1}$ & $\mathbf{1}$ & $\mathbf{1}$ & $\mathbf{1}$ & $\mathbf{1}$ & $%
\mathbf{1}$ & $\mathbf{1}$ & $\mathbf{1}$ & $\mathbf{1}$ & $\mathbf{1}$ \\ 
\hline
$SU_{2L}$ & $\mathbf{2}$ & $\mathbf{2}$ & $\mathbf{1}$ & $\mathbf{1}$ & $%
\mathbf{1}$ & $\mathbf{1}$ & $\mathbf{1}$ & $\mathbf{1}$ & $\mathbf{1}$ & $%
\mathbf{1}$ & $\mathbf{1}$ & $\mathbf{1}$ & $\mathbf{1}$ & $\mathbf{1}$ \\ 
\hline
$U_{1Y}$ & $\frac{1}{2}$ & $\frac{1}{2}$ & $0$ & $0$ & $0$ & $0$ & $0$ & $0$
& $0$ & $1$ & $1$ & $1$ & $1$ & $1$ \\ \hline
$U_{1X}$ & $1$ & $2$ & $-1$ & $-1$ & $-2$ & $0$ & $0$ & $0$ & $1$ & $5$ & $2$
& $3$ & $2$ & $3$ \\ \hline
$Z_{2}^{(1)}$ & $1$ & $1$ & $1$ & $1$ & $-1$ & $1$ & $-1$ & $-1$ & $-1$ & $%
-1 $ & $1$ & $1$ & $-1$ & $-1$ \\ \hline
$Z_{2}^{(2)}$ & $1$ & $-1$ & $1$ & $-1$ & $-1$ & $-1$ & $-1$ & $1$ & $-1$ & $%
1$ & $1$ & $-1$ & $1$ & $1$ \\ \hline
\end{tabular}%
\caption{Scalars assignments under the $SU_{3c}\times SU_{2L}\times
U_{1Y}\times U_{1X}\times Z_{2}^{(1)}\times Z_{2}^{(2)}$ symmetry.}
\label{tab:scalars}
\end{table}
\begin{table}[th]
\centering%
\begin{tabular}{|c||c|c|c|c|c|c|c|c|c|c|c|c|c|c|c|c|c|c|c|c|c|}
\hline\hline
Field & $q_{1L}$ & $q_{2L}$ & $q_{3L}$ & $u_{1R}$ & $u_{2R}$ & $u_{3R}$ & $%
d_{1R}$ & $d_{2R}$ & $d_{3R}$ & $T_{L}$ & $T_{R}$ & $\widetilde{T}_{L}$ & $%
\widetilde{T}_{R}$ & $B_{1L}$ & $B_{1R}$ & $B_{2L}$ & $B_{2R}$ & $B_{3L}$ & $%
B_{3R}$ & $B_{4L}$ & $B_{4R}$ \\ \hline
$SU_{3c}$ & $\mathbf{3}$ & $\mathbf{3}$ & $\mathbf{3}$ & $\mathbf{3}$ & $%
\mathbf{3}$ & $\mathbf{3}$ & $\mathbf{3}$ & $\mathbf{3}$ & $\mathbf{3}$ & $%
\mathbf{3}$ & $\mathbf{3}$ & $\mathbf{3}$ & $\mathbf{3}$ & $\mathbf{3}$ & $%
\mathbf{3}$ & $\mathbf{3}$ & $\mathbf{3}$ & $\mathbf{3}$ & $\mathbf{3}$ & $%
\mathbf{3}$ & $\mathbf{3}$ \\ \hline
$SU_{2L}$ & $\mathbf{2}$ & $\mathbf{2}$ & $\mathbf{2}$ & $\mathbf{1}$ & $%
\mathbf{1}$ & $\mathbf{1}$ & $\mathbf{1}$ & $\mathbf{1}$ & $\mathbf{1}$ & $%
\mathbf{1}$ & $\mathbf{1}$ & $\mathbf{1}$ & $\mathbf{1}$ & $\mathbf{1}$ & $%
\mathbf{1}$ & $\mathbf{1}$ & $\mathbf{1}$ & $\mathbf{1}$ & $\mathbf{1}$ & $%
\mathbf{1}$ & $\mathbf{1}$ \\ \hline
$U_{1Y}$ & $\frac{1}{6}$ & $\frac{1}{6}$ & $\frac{1}{6}$ & $\frac{2}{3}$ & $%
\frac{2}{3}$ & $\frac{2}{3}$ & $-\frac{1}{3}$ & $-\frac{1}{3}$ & $-\frac{1}{3%
}$ & $\frac{2}{3}$ & $\frac{2}{3}$ & $\frac{2}{3}$ & $\frac{2}{3}$ & $-\frac{%
1}{3}$ & $-\frac{1}{3}$ & $-\frac{1}{3}$ & $-\frac{1}{3}$ & $-\frac{1}{3}$ & 
$-\frac{1}{3}$ & $-\frac{1}{3}$ & $-\frac{1}{3}$ \\ \hline
$U_{1X}$ & $0$ & $0$ & $1$ & $2$ & $2$ & $2$ & $-1$ & $-1$ & $-1$ & $1$ & $2$
& $1$ & $1$ & $0$ & $-1$ & $0$ & $-1$ & $-2$ & $-2$ & $-3$ & $-3$ \\ \hline
$Z_{2}^{(1)}$ & $1$ & $1$ & $1$ & $-1$ & $-1$ & $1$ & $-1$ & $-1$ & $-1$ & $%
1 $ & $1$ & $-1$ & $-1$ & $1$ & $1$ & $1$ & $1$ & $1$ & $1$ & $1$ & $1$ \\ 
\hline
$Z_{2}^{(2)}$ & $-1$ & $-1$ & $-1$ & $-1$ & $-1$ & $-1$ & $-1$ & $-1$ & $-1$
& $1$ & $1$ & $-1$ & $-1$ & $1$ & $1$ & $1$ & $1$ & $1$ & $1$ & $-1$ & $-1$
\\ \hline
\end{tabular}%
\caption{Quark assignments under the $SU_{3c}\times SU_{2L}\times
U_{1Y}\times U_{1X}\times Z_{2}^{(1)}\times Z_{2}^{(2)}$ symmetry.}
\label{tab:quarks}
\end{table}

\begin{table}[th]
\centering%
\begin{tabular}{|c||c|c|c|c|c|c|c|c|c|c|c|c|c|c|c|c|c|c|}
\hline\hline
Field & $l_{1L}$ & $l_{2L}$ & $l_{3L}$ & $l_{1R}$ & $l_{2R}$ & $l_{3R}$ & $%
E_{1L}$ & $E_{1R}$ & $E_{2L}$ & $E_{2R}$ & $E_{3L}$ & $E_{3R}$ & $\nu _{1R}$
& $\nu _{2R}$ & $\nu _{3R}$ & $\Omega _{1R}$ & $\Omega _{2R}$ & $\Psi _{R}$
\\ \hline
$SU_{3c}$ & $\mathbf{1}$ & $\mathbf{1}$ & $\mathbf{1}$ & $\mathbf{1}$ & $%
\mathbf{1}$ & $\mathbf{1}$ & $\mathbf{1}$ & $\mathbf{1}$ & $\mathbf{1}$ & $%
\mathbf{1}$ & $\mathbf{1}$ & $\mathbf{1}$ & $\mathbf{1}$ & $\mathbf{1}$ & $%
\mathbf{1}$ & $\mathbf{1}$ & $\mathbf{1}$ & $\mathbf{1}$ \\ \hline
$SU_{2L}$ & $\mathbf{2}$ & $\mathbf{2}$ & $\mathbf{2}$ & $\mathbf{1}$ & $%
\mathbf{1}$ & $\mathbf{1}$ & $\mathbf{1}$ & $\mathbf{1}$ & $\mathbf{1}$ & $%
\mathbf{1}$ & $\mathbf{1}$ & $\mathbf{1}$ & $\mathbf{1}$ & $\mathbf{1}$ & $%
\mathbf{1}$ & $\mathbf{1}$ & $\mathbf{1}$ & $\mathbf{1}$ \\ \hline
$U_{1Y}$ & $-\frac{1}{2}$ & $-\frac{1}{2}$ & $-\frac{1}{2}$ & $-1$ & $-1$ & $%
-1$ & $-1$ & $-1$ & $-1$ & $-1$ & $-1$ & $-1$ & $0$ & $0$ & $0$ & $0$ & $0$
& $0$ \\ \hline
$U_{1X}$ & $0$ & $-3$ & $0$ & $-3$ & $-6$ & $-3$ & $-3$ & $-2$ & $-6$ & $-5$
& $-3$ & $-2$ & $2$ & $-1$ & $2$ & $-1$ & $1$ & $0$ \\ \hline
$Z_{2}^{(1)}$ & $1$ & $-1$ & $1$ & $1$ & $-1$ & $1$ & $-1$ & $-1$ & $-1$ & $%
-1$ & $1$ & $1$ & $1$ & $-1$ & $1$ & $-1$ & $-1$ & $1$ \\ \hline
$Z_{2}^{(2)}$ & $-1$ & $-1$ & $-1$ & $-1$ & $-1$ & $-1$ & $1$ & $1$ & $1$ & $%
1$ & $1$ & $1$ & $1$ & $1$ & $1$ & $-1$ & $1$ & $1$ \\ \hline
\end{tabular}%
\caption{Lepton charge assignments under the $SU_{3c}\times SU_{2L}\times
U_{1Y}\times U_{1X}\times Z_{2}^{(1)}\times Z_{2}^{(2)}$ symmetry.}
\label{tab:leptons}
\end{table}

Let us note that the SM-type Higgs doublet, i.e., $\phi _{1}$, as well as
the electroweak-singlet scalars $\sigma _{1}$ and $\rho _{3}$ are the only
scalar fields neutral under the exact $Z_{2}^{\left( 2\right) }$ discrete
symmetry. Since the $Z_{2}^{\left( 2\right) }$ symmetry is preserved, the
Higgs doublet $\phi _{1}$ and the singlets $\sigma _{1}$ and $\rho _{3}$ are
the only scalar fields that acquire nonvanishing VEVs. The VEV in $\sigma
_{1}$ is required to spontaneously break the $U_{1X}$ local symmetry,
whereas the $\rho _{3}$ VEV spontaneously breaks the $Z_{2}^{\left( 1\right)
}$ discrete symmetry, due to its nontrivial $Z_{2}^{\left( 1\right) }$
charge.

Note that the exact $Z_{2}^{\left( 2\right) }$ discrete symmetry guarantees
the presence of several stable scalar dark matter candidates in our model.
These are represented by the neutral components of the inert $SU_{2L}$
scalar doublet $\phi _{2}$, as well as by the real and imaginary parts of
the SM-singlet scalars $\sigma _{2}$, $\sigma _{3}$, $\rho _{1}$, $\rho _{2}$
and $\eta $. Furthermore, the model can have a fermionic DM candidate, which
is the only SM-singlet Majorana neutrino $\Omega _{1R}$ with a non-trivial $%
Z_{2}^{\left( 2\right) }$ charge. 

The set of $SU_{2L}$-singlet heavy quarks $T_{L}$, $T_{R}$, $B_{iL}$, $%
B_{iR} $ ($i=1,2,3$) represents the minimal amount of exotic quark degrees
of freedom needed to implement the one-loop radiative seesaw mechanism that
gives rise to the charm, bottom and strange quark masses. Furthermore, in
order to ensure the radiative seesaw mechanism responsible for the
generation of the up and down quark masses at two-loop level, the $SU_{2L}$
singlet heavy quarks $\widetilde{T}_{L}$, $\widetilde{T}_{R}$, $B_{4L}$, $%
B_{4R}$, as well as the electrically neutral, $\sigma _{3}$, $\rho _{2}$,
and electrically charged, $\varphi _{1}^{+}$, $\varphi _{2}^{+}$ scalar $%
SU_{2L}$-singlets should also be present in the particle spectrum.

To summarize, the SM fermion sector of the considered model includes a total
of six electrically charged weak-singlet leptons $E_{jL}$ and $E_{jR}$ ($%
j=1,2,3$), four right-handed neutrinos $\nu _{jR}$ ($j=1,2,3$), $\Omega _{R}$%
, and twelve $SU_{2L}$-singlet heavy quarks $T_{L}$, $T_{R}$, $\widetilde{T}%
_{L}$,$\widetilde{T}_{R}$, $B_{kL}$, $B_{kR}$ ($k=1,2,3,4$). It is assumed
that the heavy exotic $T$, $\widetilde{T}$ and $B_{k}$ quarks have electric
charges equal to $2/3$ and $-1/3$, respectively.

More specifically, the quark sector of the extended IDM under consideration
has the following $SU_{3c}\times SU_{2L}\times U_{1Y}\times U_{1X}$ charges 
\begin{eqnarray}
q_{nL} &\sim &\left( \mathbf{3,2},\frac{1}{6},0\right) ,\hspace{1cm}\hspace{%
1cm}q_{3L}\sim \left( \mathbf{3,2},\frac{1}{6},1\right) ,\hspace{1cm}\hspace{%
1cm}n=1,2,  \notag \\
u_{jR} &\sim &\left( \mathbf{3,1},\frac{2}{3},2\right) ,\hspace{1cm}\hspace{%
1cm}d_{jR}\sim \left( \mathbf{3,1},-\frac{1}{3},-1\right) ,\hspace{1cm}%
\hspace{1cm}j=1,2,3,  \notag \\
T_{L} &\sim &\left( \mathbf{3,1},\frac{2}{3},1\right) ,\hspace{1cm}\hspace{%
1cm}T_{R}\sim \left( \mathbf{3,1},\frac{2}{3},2\right) ,\hspace{1cm}  \notag
\\
\widetilde{T}_{L} &\sim &\left( \mathbf{3,1},\frac{2}{3},1\right) ,\hspace{%
1cm}\hspace{1cm}\widetilde{T}_{R}\sim \left( \mathbf{3,1},\frac{2}{3}%
,1\right) ,  \notag \\
B_{nL} &\sim &\left( \mathbf{3,1},-\frac{1}{3},0\right) ,\hspace{1cm}\hspace{%
1cm}B_{nR}\sim \left( \mathbf{3,1},-\frac{1}{3},-1\right) ,\hspace{1cm}%
\hspace{1cm}  \notag \\
B_{3L} &\sim &\left( \mathbf{3,1},-\frac{1}{3},-2\right) \,,\hspace{1cm}%
\hspace{1cm}B_{3R}\sim \left( \mathbf{3,1},-\frac{1}{3},-2\right) \,,  \notag
\\
B_{4L} &\sim &\left( \mathbf{3,1},-\frac{1}{3},-3\right) \,,\hspace{1cm}%
\hspace{1cm}B_{4R}\sim \left( \mathbf{3,1},-\frac{1}{3},-3\right)
\end{eqnarray}%
while their $Z_{2}^{\left( 1\right) }\times Z_{2}^{\left( 2\right) }$ charge
assignments read 
\begin{eqnarray}
q_{nL} &\sim &\left( 1,-1\right) ,\hspace{1cm}q_{3L}\sim \left( 1,-1\right) ,%
\hspace{1cm}u_{nR}\sim \left( -1,-1\right) ,\hspace{1cm}u_{3R}\sim \left(
1,-1\right) ,\hspace{1cm}d_{jR}\sim \left( -1,-1\right) ,  \notag \\
T_{L} &\sim &\left( 1,1\right) ,\hspace{0.8cm}T_{R}\sim \left( 1,1\right) ,%
\hspace{0.8cm}\widetilde{T}_{L}\sim \left( -1,-1\right) ,\hspace{0.8cm}%
\widetilde{T}_{R}\sim \left( -1,-1\right) ,\hspace{0.8cm}n=1,2,\hspace{1cm}%
j=1,2,3,  \notag \\
B_{jL} &\sim &\left( 1,1\right) ,\hspace{0.8cm}B_{jR}\sim \left( 1,1\right) ,%
\hspace{0.8cm}B_{4L}\sim \left( 1,-1\right) ,\hspace{1cm}B_{4R}\sim \left(
1,-1\right) \,.
\end{eqnarray}
A summary of all the field assignments with respect to the model symmetry is given in Tables \ref{tab:scalars}, \ref{tab:quarks} and \ref{tab:leptons}.

The radiative seesaw mechanism that generates the charged lepton mass
hierarchy is similar to the one that produces the SM down-type quark mass
hierarchy. The generation of one-loop tau and muon masses is mediated by the
electrically charged weak-singlet leptons $E_{rL}$ and $E_{rR}$ ($r=2,3$),
by the inert scalar $SU_{2L}$-doublet, $\phi _{2}$, and by the $SU_{2L}$%
-singlets $\sigma _{2}$, $\rho _{1}$. On the other hand, the radiative
seesaw mechanism that give rises to a two loop level electron mass is
mediated by electrically charged scalars as well as by the right-handed
Majorana neutrinos $\Psi $, $\nu _{kR}$ ($k=1,3$) and the weak-singlet
electrically charged leptons $E_{1L}$ and $E_{1R}$.

Moreover, the three-loop radiative seesaw mechanism responsible for the
generation of the light active neutrino masses is mediated by the
right-handed neutrinos $\nu _{jR}$ ($j=1,2,3$), $\Omega _{R}$, as well as by
the inert scalar $SU_{2L}$ doublet $\phi _{2}$ and the $SU_{2L}$-singlet $%
\sigma _{2}$. To avoid tree-level mixing between the right-handed Majorana
neutrinos $\nu _{kR}$ ($k=1,3$) and $\nu _{2R}$ triggered by Yukawa
interactions with $\sigma _{1}$, we need to impose a nontrivial $%
Z_{2}^{\left( 1\right) }$ charge of $\nu _{2R}$ while keeping $\nu _{kR}$ ($%
k=1,3$) $Z_{2}^{\left( 1\right) }$-neutral.

In particular, the $SU_{3c}\times SU_{2L}\times U_{1Y}\times U_{1X}$ charges
of the leptonic and neutrino fields of the model are defined as follows 
\begin{eqnarray}
l_{kL} &\sim &\left( \mathbf{1,2},-\frac{1}{2},0\right) ,\hspace{1cm}\hspace{%
1cm}l_{2L}\sim \left( \mathbf{1,2},-\frac{1}{2},-3\right) ,\hspace{1cm}k=1,3,
\notag \\
l_{kR} &\sim &\left( \mathbf{1,1},-1,-3\right) ,\hspace{1cm}\hspace{1cm}%
l_{2R}\sim \left( \mathbf{1,1},-1,-6\right) ,  \notag \\
E_{1L} &\sim &\left( \mathbf{1,1},-1,-3\right) ,\hspace{1cm}\hspace{1cm}%
E_{1R}\sim \left( \mathbf{1,1},-1,-2\right) ,  \notag \\
E_{2L} &\sim &\left( \mathbf{1,1},-1,-6\right) ,\hspace{1cm}\hspace{1cm}%
E_{2R}\sim \left( \mathbf{1,1},-1,-5\right) ,  \notag \\
E_{3L} &\sim &\left( \mathbf{1,1},-1,-3\right) ,\hspace{1cm}\hspace{1cm}%
E_{3R}\sim \left( \mathbf{1,1},-1,-2\right) ,  \notag \\
\nu _{kR} &\sim &\left( \mathbf{1,1},0,2\right) ,\hspace{1cm}\hspace{1cm}%
\hspace{1cm}\nu _{2R}\sim \left( \mathbf{1,1},0,-1\right) ,\hspace{1cm} 
\notag \\
\Omega _{1R} &\sim &\left( \mathbf{1,1},0,-1\right) ,\hspace{1cm}\Omega
_{2R}\sim \left( \mathbf{1,1},0,1\right) ,\hspace{1cm}\Psi _{R}\sim \left( 
\mathbf{1,1},0,0\right) \,,
\end{eqnarray}%
whereas the corresponding $Z_{2}^{\left( 1\right) }\times Z_{2}^{\left(
2\right) }$ charges are given by 
\begin{eqnarray}
l_{kL} &\sim &\left( 1,-1\right) ,\hspace{0.5cm}l_{2L}\sim \left(
-1,-1\right) ,\hspace{0.5cm}l_{kR}\sim \left( 1,-1\right) ,\hspace{0.5cm}%
l_{2R}\sim \left( -1,-1\right) ,\hspace{0.5cm}k=1,3,  \notag \\
E_{1L} &\sim &\left( -1,1\right) ,\hspace{0.5cm}E_{1R}\sim \left(
-1,1\right) ,\hspace{0.5cm}E_{2L}\sim \left( -1,1\right) ,\hspace{0.5cm}%
E_{2R}\sim \left( -1,1\right) ,\hspace{0.5cm}E_{3L}\sim \left( 1,1\right) ,%
\hspace{0.5cm}E_{3R}\sim \left( 1,1\right) ,  \notag \\
\nu _{kR} &\sim &\left( 1,1\right) ,\hspace{0.5cm}\nu _{2R}\sim \left(
-1,1\right) ,\hspace{1cm}\Omega _{1R}\sim \left( -1,-1\right) ,\hspace{0.5cm}%
\Omega _{2R}\sim \left( -1,1\right) ,\hspace{0.5cm}\Psi _{R}\sim \left(
1,1\right) ,\hspace{0.5cm}k=1,3\,.
\end{eqnarray}%
The $U_{1X}\times Z_{2}^{\left( 1\right) }\times Z_{2}^{\left( 2\right) }$
symmetry and the particular assignments listed above are crucial for
avoiding the appearance of SM light active neutrino masses at one- and
two-loop levels.

Let us note that the left-handed quark $SU_{2L}$ doublets of the first and
second generations are distinguished from the third generation by means of
the $U_{1X}$ charge assignments. In addition, the local $U_{1X}$ family
symmetry distinguishes the second generation left-handed lepton $SU_{2L}$
doublet from the first and third generation ones. Such non-universal $U_{1X}$
charge assignments in the fermion sector are crucial for implementing the
sequential loop suppression mechanism and, hence, for the induced strong
hierarchies in the SM fermion mass spectrum. Besides, as will be explicitly
demonstrated in a forthcoming paper, such assignments are also relevant for
explaining the $R_{K}$ and $R_{K^{\ast}}$ anomalies.

With the above assignments, we have numerically checked that the gauge
anomaly cancellation conditions 
\begin{eqnarray}
A_{\left[ SU_{3c}\right] ^{2}U_{1X}}
&=&\sum_{Q}X_{Q_{L}}-\sum_{Q}X_{Q_{R}}\,,\qquad A_{\left[ SU_{2L}\right]
^{2}U_{1X}}=\sum_{L}X_{L_{L}}+3\sum_{Q}X_{Q_{L}},  \notag \\
A_{\left[ U_{1Y}\right] ^{2}U_{1X}} &=&\sum_{L,Q}\left(
Y_{L_{L}}^{2}X_{L_{L}}+3Y_{Q_{L}}^{2}X_{Q_{L}}\right) -\sum_{L,Q}\left(
Y_{L_{R}}^{2}X_{L_{R}}+3Y_{Q_{R}}^{2}X_{Q_{R}}\right) ,  \notag \\
A_{\left[ U_{1X}\right] ^{2}U_{1Y}} &=&\sum_{L,Q}\left(
Y_{L_{L}}X_{L_{L}}^{2}+3Y_{Q_{L}}X_{Q_{L}}^{2}\right) -\sum_{L,Q}\left(
Y_{L_{R}}X_{L_{R}}^{2}+3Y_{Q_{R}}X_{Q_{R}}^{2}\right) ,  \notag \\
A_{\left[ U_{1X}\right] ^{3}} &=&\sum_{L,Q}\left(
X_{L_{L}}^{3}+3X_{Q_{L}}^{3}\right) -\sum_{L,Q}\left(
X_{L_{R}}^{3}+X_{N_{R}}^{3}+3X_{Q_{R}}^{3}\right) ,  \notag \\
A_{\left[ \mathrm{Gravity}\right] ^{2}U_{1X}} &=&\sum_{L,Q}\left(
X_{L_{L}}+3X_{Q_{L}}\right) -\sum_{L,Q,N}\left(
X_{L_{R}}+X_{N_{R}}+3X_{Q_{R}}\right)
\end{eqnarray}%
are satisfied in our model. Let us note that in the expression for $A_{\left[
SU_{2L}\right] ^{2}U_{1X}}$ the sum is performed only over the $SU_{2L}$
doublets of left handed fermionic fields. On the other hand, in the
expression for $A_{\left[ \mathrm{Gravity}\right] ^{2}U_{1X}}$, the sum is
performed over all left handed fermionic fields.


\subsection{Yukawa interactions}

\label{Sec:Yukawa} 

With the above specified particle content and charge assignments, the most
general renormalizable Lagrangian of Yukawa interactions and the exotic
fermion mass terms, invariant under the $SU_{3c}\times SU_{2L}\times
U_{1Y}\times U_{1X}\times Z_{2}^{\left( 1\right) }\times Z_{2}^{\left(
2\right) }$ symmetry, takes the following form

\begin{eqnarray}
\tciLaplace _{\mathrm{F}} &=&y_{3j}^{\left( u\right) }\overline{q}_{3L}%
\widetilde{\phi }_{1}u_{3R}+\sum_{n=1}^{2}x_{n}^{\left( u\right) }\overline{q%
}_{nL}\widetilde{\phi }_{2}T_{R}+\sum_{n=1}^{2}z_{j}^{\left( u\right) }%
\overline{T}_{L}\eta ^{\ast }u_{nR}+y_{T}\overline{T}_{L}\sigma _{1}T_{R}+m_{%
\widetilde{T}}\overline{\widetilde{T}}_{L}\widetilde{T}_{R}+x^{\left(
T\right) }\overline{T}_{L}\rho _{2}\widetilde{T}_{R}  \notag \\
&&+\sum_{n=1}^{2}x_{n}^{\left( d\right) }\overline{q}_{3L}\phi
_{2}B_{nR}+\sum_{n=1}^{2}\sum_{j=1}^{3}y_{nj}^{\left( d\right) }\overline{B}%
_{nL}\eta d_{jR}+\sum_{j=1}^{3}z_{j}^{\left( d\right) }\overline{B}_{3L}\eta
^{\ast }d_{jR}+\sum_{n=1}^{2}w_{n}^{\left( u\right) }\overline{B}%
_{4L}\varphi _{1}^{-}u_{nR}  \notag \\
&&+\sum_{k=3}^{4}m_{B_{k}}\overline{B}_{kL}B_{kR}+\sum_{n=1}^{2}x_{n}^{%
\left( d\right) }\overline{q}_{nL}\phi
_{2}B_{3R}+\sum_{n=1}^{2}\sum_{m=1}^{2}y_{nm}^{\left( B\right) }\overline{B}%
_{nL}\sigma _{1}^{\ast }B_{mR}+z^{\left( B\right) }\overline{B}_{3L}\sigma
_{2}^{\ast }B_{4R}+\sum_{j=1}^{3}w_{j}^{\left( d\right) }\overline{%
\widetilde{T}}_{L}\varphi _{2}^{+}d_{jR}  \notag \\
&&+\sum_{k=1,3}x_{k3}^{\left( l\right) }\overline{l}_{kL}\phi
_{2}E_{3R}+\sum_{k=1,3}y_{3k}^{\left( l\right) }\overline{E}_{3L}\rho
_{1}l_{kR}+x_{22}^{\left( l\right) }\overline{l}_{2L}\phi
_{2}E_{2R}+y_{22}^{\left( l\right) }\overline{E}_{2L}\rho _{1}l_{2R}  \notag
\\
&&+\sum_{i=1}^{3}y_{i}^{\left( E\right) }\overline{E}_{iL}\sigma _{1}^{\ast
}E_{iR}+x_{2}^{\left( \nu \right) }\overline{l}_{2L}\widetilde{\phi }_{2}\nu
_{2R}+\sum_{k=1,3}z_{k}^{\left( l\right) }\overline{\Psi _{R}^{C}}\varphi
_{3}^{+}l_{kR}+\sum_{k=1,3}z_{k}^{\left( \nu \right) }\overline{E}%
_{1L}\varphi _{1}^{-}\nu _{kR}+z^{\left( E\right) }\overline{\Psi _{R}^{C}}%
\varphi _{4}^{+}E_{1R}  \notag \\
&&+\sum_{k=1,3}\sum_{n=1,3}x_{kn}^{\left( \nu \right) }\overline{l}_{kL}%
\widetilde{\phi }_{2}\nu _{nR}+\sum_{k=1,3}y_{k}^{\left( \Omega \right) }%
\overline{\Omega _{1R}^{C}}\eta ^{\ast }\nu _{kR}+y^{\left( \Omega \right) }%
\overline{\Omega _{1R}^{C}}\sigma _{3}^{\ast }\nu _{2R}  \notag \\
&&+x_{1}^{\left( \Psi \right) }\overline{\Omega _{1R}^{C}}\eta \Psi
_{R}+x_{2}^{\left( \Psi \right) }\overline{\Omega _{2R}^{C}}\eta ^{\ast
}\Psi _{R}+z_{\Omega }\overline{\Omega _{1R}^{C}}\sigma _{2}^{\ast }\Omega
_{2R}+m_{\Psi }\overline{\Psi _{R}^{C}}\Psi _{R}+h.c.\,,  \label{LY}
\end{eqnarray}%
where the Yukawa couplings are $\mathcal{O}(1)$ parameters. 
From the quark Yukawa terms it follows that the top quark mass emerges due to an
interaction involving the SM-like Higgs doublet $\phi _{1}$ only. After
spontaneous breaking of the electroweak symmetry, the observed hierarchies
of SM fermion masses arise by means of a sequential loop suppression,
according to the following pattern: tree-level top quark mass; one-loop
bottom, strange, charm, tau and muon masses; two-loop masses for the up,
down quarks as well as for the electron. Furthermore, the SM light active
neutrinos get their masses by means of a three-loop radiative seesaw
mechanism.

A few comments on the phenomenological implications of the Lagrangian (\ref{LY}) are in order.
 Notice that the neutrino Yukawa coupling to, e.g. SM lepton and $\phi_2$ doublets 
$x_{kn}^{(\nu)}$, is generally not suppressed. On the other hand it can be constrained by
Z-boson decays. However, we do not expect this constraint to be very strong provided that $\phi_2$ boson is heavy. Indeed, the corresponding one-loop amplitude of Z-boson decay would be suppressed by a large mass of $\phi_2$ scalar in the propagators. Therefore, such an amplitude is, in any case, expected to be small.

It is also worth mentioning that the Yukawa interactions $\overline{E}_{2L}\rho
_{1}l_{2R}$ and $\overline{l}_{2L}\phi _{2}E_{2R}$ presented in the 4th line
of Eq.~(\ref{LY}) as well as the trilinear scalar interaction $\rho
_{2}\left( \phi _{1}\cdot \phi _{2}^{\dagger }\right) \sigma _{1}^{\ast }$
generate a one loop level scalar contribution to the muon anomalous magnetic
moment  $g-2$. The exchange with the heavy $Z^{\prime }$ gauge boson
also yields a contribution  to this observable. The possibility of explanation of the observed deviation of the $g-2$ from the SM value will be studied in the forthcoming publication.

Let us also note that
from the term $\overline{l}_{kL}\widetilde{\phi }_{2}\nu _{nR}$
in Eq. (\ref{LY}), it follows that the charged lepton flavor violating decay 
$\tau \rightarrow e\gamma $ is induced at one loop level by electrically
charged scalar $\phi _{2}^{+}$ (arising from the $SU_{2L}$ inert doublet 
$\phi $) and right handed Majorana neutrinos $\nu _{sR}$ ($s=1,3$) (whose
masses are generated at two loop level ) appearing in the internal lines of
the loop. 
This decay also receives a one
loop level contribution arising from the $Z^{\prime }$ exchange.

Due to the fact that electron is charged under $U_{1X}$ the LEP measurements of 
$e^{+}e^{-}\rightarrow \mu ^{+}\mu ^{-}$ set a stringent limit \cite{CarcamoHernandez:2019xkb} on the ratio
\begin{equation}
\frac{M_{Z^{\prime }}}{g_{X}}>12\;\mbox{TeV}\,.
\end{equation}
Our model contains two electroweak doublet Higgs scalars $\phi_{1,2}$. As such we should take special care of  Flavor Changing Neutral Currents (FCNCs).  
%
Our model automatically implements the alignment limit for the lightest 125 GeV Higgs boson, since all other
scalar states appear to be decoupled in the mass spectrum and, hence, are very heavy by default. This means
the SM-like Higgs boson state does not have tree-level FCNCs while such contributions from the heavier scalars
are strongly suppressed by their large mass scale. While a detailed study of the FCNC constraints goes beyond the scope of the current work, we can 
make a generic statement about nonexistence of FCNCs in our model based upon the Glashow-Weinberg-Paschos 
theorem  \cite{Glashow:1976nt,Paschos:1976ay}. This theorem states that there will be no tree-level FCNC coming from the scalar sector, if all right-handed 
fermions of a given electric charge couple to only one of the doublets. As seen from Eq.~(\ref{LY}) this condition is satisfied in our model.
So, despite of an obvious mass suppression, any possible FCNC corrections would emerge at a loop level only,
yielding the model safe with respect to the corresponding phenomenological constraints. Finally, any possible FCNC from the $Z'$ mediation would, for sure, be very much suppressed by its large mass 
scale compared to the EW one, i.e. $m_{Z'} >12$ TeV (for $g_X=1$), according to the LEP constraint.

\subsection{Scalar potential}
\
\label{Sec:potential} 

The most general renormalizable scalar potential invariant under the gauge
and discrete symmetries of our model is given by 
\begin{eqnarray}
V &=&\sum_{i=1}^{2}\left( \mu _{pi}^{2}\left\vert \phi _{i}\right\vert
^{2}+\lambda _{pi}\left\vert \phi _{i}\right\vert ^{4}\right)
+\sum_{j=1}^{3}\left( \mu _{sj}^{2}\left\vert \sigma _{j}\right\vert
^{2}+\lambda _{sj}\left\vert \sigma _{j}\right\vert ^{4}\right)
+\sum_{j=1}^{3}\left( \mu _{rj}^{2}\left\vert \rho _{j}\right\vert
^{2}+\lambda _{rj}\left\vert \rho _{j}\right\vert ^{4}\right) +\mu
_{e}^{2}\left\vert \eta \right\vert ^{2}+\lambda _{e}\left\vert \eta
\right\vert ^{4}  \notag \\
&&+\sum_{i=1}^{5}\left( \mu _{fi}^{2}\varphi _{i}^{+}\varphi
_{i}^{-}+\lambda _{fi}\left( \varphi _{i}^{+}\varphi _{i}^{-}\right)
^{2}\right) +\sum_{i=1}^{2}\sum_{j=1}^{3}\alpha _{ij}\left\vert \phi
_{i}\right\vert ^{2}\left\vert \sigma _{j}\right\vert
^{2}+\sum_{i=1}^{2}\sum_{j=1}^{3}\beta _{ij}\left\vert \phi _{i}\right\vert
^{2}\left\vert \rho _{j}\right\vert ^{2}+\sum_{i=1}^{2}\kappa
_{pi}\left\vert \phi _{i}\right\vert ^{2}\left\vert \eta \right\vert ^{2} 
\notag \\
&&+\kappa _{1}\left\vert \phi _{1}\right\vert ^{2}\left\vert \phi
_{2}\right\vert ^{2}+\kappa _{2}\left( \phi _{1}\phi _{2}^{\dagger }\right)
\left( \phi _{2}\phi _{1}^{\dagger }\right) +\kappa _{3}\left[ \varepsilon
_{ab}\varepsilon _{cd}\left( \phi _{1}\right) ^{a}\left( \phi _{2}\right)
^{b}\left( \phi _{1}^{\dagger }\right) ^{c}\left( \phi _{2}^{\dagger
}\right) ^{d}+h.c.\right]  \notag \\
&&+\sum_{i=1}^{3}\sum_{j=1}^{3}\gamma _{ij}\left\vert \sigma _{i}\right\vert
^{2}\left\vert \rho _{j}\right\vert ^{2}+\sum_{i=1}^{3}\alpha
_{ei}\left\vert \eta \right\vert ^{2}\left\vert \sigma _{i}\right\vert
^{2}+\sum_{j=1}^{3}\beta _{ej}\left\vert \eta \right\vert ^{2}\left\vert
\rho _{j}\right\vert ^{2}+\sum_{i=1}^{5}\sum_{j=1}^{5}\kappa _{ij}\left(
\varphi _{i}^{+}\varphi _{i}^{-}\right) \left( \varphi _{j}^{+}\varphi
_{j}^{-}\right)  \notag \\
&&+\sum_{i=1}^{5}\sum_{j=1}^{2}\lambda _{ij}\left( \varphi _{i}^{+}\varphi
_{i}^{-}\right) \left\vert \phi _{j}\right\vert
^{2}+\sum_{i=1}^{5}\sum_{j=1}^{3}\varsigma _{ij}\left( \varphi
_{i}^{+}\varphi _{i}^{-}\right) \left\vert \sigma _{j}\right\vert
^{2}+\sum_{i=1}^{5}\sum_{j=1}^{3}\varrho _{ij}\left( \varphi _{i}^{+}\varphi
_{i}^{-}\right) \left\vert \rho _{j}\right\vert ^{2}+\sum_{i=1}^{5}\varkappa
_{i}\left( \varphi _{i}^{+}\varphi _{i}^{-}\right) \left\vert \eta
\right\vert ^{2}  \notag \\
&&+A_{1}\left[ \left( \phi _{1}^{\dagger }\cdot \phi _{2}\right) \sigma
_{2}+h.c.\right] +A_{2}\left[ \varepsilon _{ab}\left( \phi _{1}\right)
^{a}\left( \phi _{2}\right) ^{b}\varphi _{3}^{-}+h.c.\right] +A_{3}\left(
\varphi _{4}^{-}\varphi _{5}^{+}\sigma _{1}+h.c.\right) +A_{4}\left( \rho
_{1}\sigma _{2}\sigma _{1}^{\ast }+h.c.\right)  \notag \\
&&+A_{5}\left( \eta \sigma _{2}\rho _{3}+h.c\right) +A_{6}\left( \rho
_{1}\rho _{2}\rho _{3}+h.c.\right) +A_{7}\left( \rho _{2}\eta \sigma
_{1}+h.c.\right) +A_{8}\left( \sigma _{3}\sigma _{1}^{\ast }\eta
+h.c.\right) +A_{9}\left( \varphi _{2}^{-}\varphi _{4}^{+}\rho
_{3}+h.c.\right)  \notag \\
&&+A_{10}\left( \varphi _{1}^{-}\varphi _{3}^{+}\sigma _{3}^{\ast
}+h.c.\right) +A_{11}\left( \varphi _{2}^{-}\varphi _{3}^{+}\sigma
_{2}+h.c.\right) +A_{11}\left( \varphi _{3}^{-}\varphi _{4}^{+}\eta
+h.c.\right) +A_{12}\left( \varphi _{3}^{-}\varphi _{5}^{+}\rho
_{2}+h.c.\right)  \notag \\
&&+\zeta _{1}\left[ \eta \left( \phi _{1}\cdot \phi _{2}^{\dagger }\right)
\rho _{3}+h.c.\right] +\zeta _{2}\left[ \varepsilon _{ab}\left( \phi
_{1}\right) ^{a}\left( \phi _{2}\right) ^{b}\varphi _{2}^{-}\sigma _{2}+h.c.%
\right] +\zeta _{3}\left[ \varepsilon _{ab}\left( \phi _{1}\right)
^{a}\left( \phi _{2}\right) ^{b}\varphi _{1}^{-}\sigma _{3}^{\ast }+h.c.%
\right]  \notag \\
&&+\zeta _{4}\left( \varphi _{1}^{+}\varphi _{5}^{-}\sigma
_{1}^{2}+h.c.\right) +\zeta _{5}\left( \sigma _{1}\sigma _{2}\sigma
_{3}^{\ast }\rho _{3}+h.c.\right) +\zeta _{6}\left( \sigma _{1}\sigma
_{2}^{\ast }\rho _{2}\rho _{3}+h.c.\right) +\zeta _{7}\left( \sigma
_{1}^{\ast }\sigma _{2}\rho _{2}\rho _{3}+h.c.\right)  \notag \\
&&+\zeta _{8}\left[ \sigma _{1}^{2}\left( \sigma _{2}^{\ast }\right)
^{2}+h.c.\right] +\zeta _{9}\left[ \sigma _{3}\rho _{2}\left( \sigma
_{1}^{\ast }\right) ^{2}+h.c.\right] +\zeta _{10}\left( \rho _{1}\eta \rho
_{3}\sigma _{1}+h.c.\right) +\zeta _{11}\left[ \rho _{1}\left( \phi
_{1}\cdot \phi _{2}^{\dagger }\right) \sigma _{1}^{\ast }+h.c.\right]  \notag
\\
&&+\zeta _{12}\left[ \rho _{1}^{\ast }\left( \phi _{1}\cdot \phi
_{2}^{\dagger }\right) \sigma _{1}^{\ast }+h.c.\right] +\zeta _{13}\left(
\varphi _{2}^{-}\varphi _{5}^{+}\sigma _{1}\rho _{3}+h.c.\right) .
\end{eqnarray}%
From the minimization conditions for this potential, we find the following
simple relations 
\begin{eqnarray}
\mu _{p1}^{2} &=&\frac{1}{2}\left( -2v^{2}\lambda _{p1}-\beta _{13}v_{\rho
}^{2}-\alpha _{11}v_{\sigma }^{2}\right) ,  \notag \\
\mu _{s1}^{2} &=&\frac{1}{2}\left( -\gamma _{13}v_{\rho }^{2}-2v_{\sigma
}^{2}\lambda _{s1}-\alpha _{11}v^{2}\right) ,  \notag \\
\mu _{r5}^{2} &=&\frac{1}{2}\left( -2\lambda _{r3}v_{\rho }^{2}-\gamma
_{13}v_{\sigma }^{2}-\beta _{13}v^{2}\right) ,
\end{eqnarray}%
that will be used below in a discussion of the scalar mass spectrum of the
model.


\section{Scalar mass spectrum}

\label{Sec:scalar-mass} 

Considering the scalar potential given above, we find that the squared mass
matrices for the CP-even neutral scalar sector are have the following form 
\begin{equation}
M_{\mathrm{CPeven}}=\left( 
\begin{array}{cc}
M_{\mathrm{CPeven}}^{\left( 1\right) } & 0_{3\times 8} \\ 
0_{8\times 3} & M_{\mathrm{CPeven}}^{\left( 2\right) }%
\end{array}%
\right) \,,  \label{MCPeven}
\end{equation}%
where $M_{\mathrm{CPeven}}^{\left( 1\right) }$ and $M_{\mathrm{CPeven}%
}^{\left( 2\right) }$ are the squared mass matrices for the $Z_{2}^{\left(
2\right) }$-neutral and $Z_{2}^{\left( 2\right) }$-charged scalars,
respectively. The matrix $M_{\mathrm{CPeven}}^{\left( 1\right) }$ in the
basis $\left( \func{Re}\left( \phi _{1}^{0}\right) ,\func{Re}\left( \sigma
_{1}\right) ,\rho _{3}\right) $ (remind, $\rho _{3}$ is a real SM-singlet
scalar), takes the form: 
\begin{equation}
M_{\mathrm{CPeven}}^{\left( 1\right) }=\left( 
\begin{array}{ccc}
v^{2}\lambda _{p1} & \frac{1}{2}vv_{\sigma }\alpha _{11} & \frac{1}{2}%
vv_{\rho }\beta _{13} \\ 
\frac{1}{2}vv_{\sigma }\alpha _{11} & v_{\sigma }^{2}\lambda _{s1} & \frac{1%
}{2}v_{\rho }v_{\sigma }\gamma _{13} \\ 
\frac{1}{2}vv_{\rho }\beta _{13} & \frac{1}{2}v_{\rho }v_{\sigma }\gamma
_{13} & v_{\rho }^{2}\lambda _{r3}%
\end{array}%
\right) \,.
\end{equation}%
The second mass form $M_{\mathrm{CPeven}}^{\left( 2\right) }$ in the basis $%
\left( \func{Re}\left( \sigma _{2}\right) ,\func{Re}\left( \sigma
_{3}\right) ,\func{Re}\left( \rho _{1}\right) ,\func{Re}\left( \rho
_{2}\right) ,\func{Re}\left( \eta \right) ,\func{Re}\left( \phi
_{2}^{0}\right) \right) $ reads 
\begin{eqnarray}
&&M_{\mathrm{CPeven}}^{\left( 2\right) }=\left( 
\begin{array}{cc}
C_{1} & C_{2} \\ 
C_{2}^{T} & C_{3}%
\end{array}%
\right) ,  \notag \\
&&C_{1}=\left( 
\begin{array}{ccc}
\frac{1}{2}\left( \frac{\alpha _{12}v^{2}}{2}+\mu _{\text{s2}}^{2}+\frac{1}{2%
}v_{\rho }^{2}\gamma _{23}+v_{\sigma }^{2}\zeta _{8}\right) & 0 & \frac{%
A_{4}v_{\sigma }}{2\sqrt{2}} \\ 
0 & \frac{1}{4}\left( \alpha _{13}v^{2}+2\mu _{\text{s3}}^{2}+v_{\rho
}^{2}\gamma _{33}\right) & 0 \\ 
\frac{A_{4}v_{\sigma }}{2\sqrt{2}} & 0 & \frac{1}{4}\left( \beta
_{12}v^{2}+2\mu _{\text{r2}}^{2}+v_{\sigma }^{2}\gamma _{11}\right)%
\end{array}%
\right) ,  \notag \\
&&C_{2}=\left( 
\begin{array}{ccc}
\frac{1}{4}v_{\rho }v_{\sigma }\left( \zeta _{6}+\zeta _{7}\right) & \frac{%
\sqrt{2}}{4}v_{\rho }A_{5} & \frac{vA_{1}}{2\sqrt{2}} \\ 
\frac{1}{4}v_{\sigma }^{2}\zeta _{9} & \frac{A_{8}v_{\sigma }}{2\sqrt{2}} & 0
\\ 
\frac{A_{6}v_{\rho }}{2\sqrt{2}} & \frac{1}{4}v_{\rho }v_{\sigma }\zeta _{10}
& \frac{1}{4}vv_{\sigma }\left( \zeta _{11}+\zeta _{12}\right)%
\end{array}%
\right) ,\hspace{1cm}\hspace{1cm} \\
&&C_{3}=\left( 
\begin{array}{ccc}
\frac{1}{4}\left( \beta _{14}v^{2}+2\mu _{\text{r4}}^{2}+v_{\sigma
}^{2}\gamma _{12}\right) & \frac{A_{7}v_{\sigma }}{2\sqrt{2}} & 0 \\ 
\frac{A_{7}v_{\sigma }}{2\sqrt{2}} & \frac{1}{4}\left( \kappa _{\text{p1}%
}v^{2}+2\mu _{e}^{2}+v_{\sigma }^{2}\alpha _{\text{e1}}+v_{\rho }^{2}\beta _{%
\text{e3}}\right) & 0 \\ 
0 & 0 & \frac{1}{4}\left( \kappa _{1}v^{2}+\kappa _{2}v^{2}+2\mu _{\text{p2}%
}^{2}+v_{\sigma }^{2}\alpha _{21}+v_{\rho }^{2}\beta _{23}\right)%
\end{array}%
\right) .  \notag
\end{eqnarray}

Since the $126$ GeV SM-like Higgs boson is found in the squared mass matrix $%
M_{\mathrm{CPeven}}^{\left( 1\right) }$, and considering the fact that the
scalar potential has a very large number of parameters, in this first study
it is sufficient to diagonalize only $M_{\mathrm{CPeven}}^{\left( 1\right) }$
in the scalar sector. In addition, since this matrix cannot be diagonalized
in analytically closed form, and for the sake of simplicity, here we focus
on a particular scenario with $v_{\sigma }=v_{\rho }$. In this scenario, the
matrix $M_{\mathrm{CPeven}}^{\left( 1\right) }$ can be diagonalized as
follows 
\begin{eqnarray}
\left( R_{\mathrm{CPeven}}^{\left( 1\right) }\right) ^{T}M_{\mathrm{CPeven}%
}^{\left( 1\right) }R_{\mathrm{CPeven}}^{\left( 1\right) } &\simeq &\left( 
\begin{array}{ccc}
\frac{8}{11}\lambda v^{2} & 0 & 0 \\ 
0 & \frac{1}{2}\left( 4-\sqrt{5}\right) \lambda v_{\sigma }^{2} & 0 \\ 
0 & 0 & \frac{1}{2}\left( 4+\sqrt{5}\right) \lambda v_{\sigma }^{2}%
\end{array}%
\right) , \\
R_{\mathrm{CPeven}}^{\left( 1\right) } &\simeq &\left( 
\begin{array}{ccc}
-1+\frac{13}{121}x^{2} & -\frac{1}{11}\sqrt{13+\frac{19}{\sqrt{5}}}x & \frac{%
1}{11}\sqrt{13-\frac{19}{\sqrt{5}}}x \\ 
\frac{5x}{11} & -\sqrt{\frac{1}{2}+\frac{1}{\sqrt{5}}} & \sqrt{\frac{1}{2}-%
\frac{1}{\sqrt{5}}} \\ 
\frac{x}{11} & \frac{1}{\sqrt{10+4\sqrt{5}}} & \sqrt{\frac{1}{2}+\frac{1}{%
\sqrt{5}}}%
\end{array}%
\right) ,\hspace{1cm}\hspace{1cm}x=\frac{v}{v_{\sigma }}.  \notag
\end{eqnarray}%
Consequently, the physical scalar states contained in the matrix $M_{\mathrm{%
CPeven}}^{\left( 1\right) }$ are given by: 
\begin{eqnarray}
\left( 
\begin{array}{c}
h \\ 
\chi _{1} \\ 
\chi _{2}%
\end{array}%
\right) &\simeq &\left( 
\begin{array}{ccc}
-1+\frac{13}{121}x^{2} & \frac{5}{11}x & \frac{1}{11}x \\ 
-\frac{1}{11}x\sqrt{\frac{19}{5}\sqrt{5}+13} & -\sqrt{\frac{1}{5}\sqrt{5}+%
\frac{1}{2}} & \frac{1}{2}\frac{\sqrt{2}}{\sqrt{2\sqrt{5}+5}} \\ 
\frac{1}{11}x\sqrt{13-\frac{19}{5}\sqrt{5}} & \sqrt{\frac{1}{2}-\frac{1}{5}%
\sqrt{5}} & \sqrt{\frac{1}{5}\sqrt{5}+\frac{1}{2}}%
\end{array}%
\right) \allowbreak \left( 
\begin{array}{c}
\phi _{1R}^{0} \\ 
\sigma _{1R} \\ 
\rho _{3}%
\end{array}%
\right) ,  \notag \\
\phi _{1R}^{0} &=&\func{Re}\left( \phi _{1}^{0}\right) ,\hspace{1cm}\hspace{%
1cm}\sigma _{1R}=\func{Re}\left( \sigma _{1}\right) ,
\end{eqnarray}%
where $h$ is the $126$ GeV SM-like Higgs boson, whereas $\chi _{1}$ and $%
\chi _{2}$ are the physical heavy scalar fields, which acquire masses at the
scale of $U_{1X}$ breaking. The squared masses of these fields are given by 
\begin{equation}
m_{h}^{2}\simeq \frac{8}{11}\lambda v^{2},\hspace{1cm}\hspace{1cm}m_{\chi
_{1}}^{2}\simeq \frac{1}{2}\left( 4-\sqrt{5}\right) \lambda v_{\sigma }^{2},%
\hspace{1cm}\hspace{1cm}m_{\chi _{2}}^{2}\simeq \frac{1}{2}\left( 4+\sqrt{5}%
\right) \lambda v_{\sigma }^{2}\,.
\end{equation}%
Furthermore, we find that the SM-like Higgs boson $h$ has the couplings that
are very close to the SM expectation, with small deviations of the order of $%
\sim v^{2}/v_{\sigma }^{2}$.

Considering the CP-odd neutral scalar sector, we find that the squared mass
matrices for the electrically neutral CP-odd scalars in the basis, are $%
\left( \func{Im}\left( \phi _{1}^{0}\right) ,\func{Im}\left( \sigma
_{1}\right) ,\func{Im}\left( \sigma _{2}\right) ,\func{Im}\left( \sigma
_{3}\right) ,\func{Im}\left( \rho _{2}\right) ,\func{Im}\left( \rho
_{4}\right) ,\func{Im}\left( \eta \right) ,\func{Im}\left( \phi
_{2}^{0}\right) ,\func{Im}\left( \rho _{1}\right) ,\func{Im}\left( \rho
_{3}\right) \right) $ are given by 
\begin{eqnarray}
&&M_{\mathrm{CPodd}}=\left( 
\begin{array}{cc}
M_{\mathrm{CPodd}}^{\left( 1\right) } & 0_{2\times 8} \\ 
0_{8\times 2} & M_{\mathrm{CPodd}}^{\left( 2\right) }%
\end{array}%
\right) ,\hspace{0.7cm}\hspace{0.7cm}M_{\mathrm{CPodd}}^{\left( 1\right)
}=0_{2\times 2},\hspace{0.7cm}\hspace{0.7cm}M_{\mathrm{CPodd}}^{\left(
2\right) }=\left( 
\begin{array}{cc}
D_{1} & D_{2} \\ 
D_{2}^{T} & D_{3}%
\end{array}%
\right) ,  \notag \\
&&D_{1}=\left( 
\begin{array}{ccc}
\frac{1}{2}\left( \frac{\alpha _{12}v^{2}}{2}+\mu _{\text{s2}}^{2}+\frac{1}{2%
}v_{\rho }^{2}\gamma _{23}-v_{\sigma }^{2}\zeta _{8}\right) & 0 & -\frac{%
A_{4}v_{\sigma }}{2\sqrt{2}} \\ 
0 & \frac{1}{4}\left( \alpha _{13}v^{2}+2\mu _{\text{s3}}^{2}+v_{\rho
}^{2}\gamma _{33}\right) & 0 \\ 
-\frac{A_{4}v_{\sigma }}{2\sqrt{2}} & 0 & \frac{1}{4}\left( \beta
_{12}v^{2}+2\mu _{\text{r2}}^{2}+v_{\sigma }^{2}\gamma _{11}\right)%
\end{array}%
\right) ,  \notag \\
&&D_{2}=\left( 
\begin{array}{ccc}
\frac{1}{4}v_{\rho }v_{\sigma }\left( \zeta _{6}-\zeta _{7}\right) & -\frac{%
\sqrt{2}}{4}v_{\rho }A_{5} & -\frac{vA_{1}}{2\sqrt{2}} \\ 
-\frac{1}{4}v_{\sigma }^{2}\zeta _{9} & -\frac{A_{8}v_{\sigma }}{2\sqrt{2}}
& 0 \\ 
-\frac{A_{6}v_{\rho }}{2\sqrt{2}} & -\frac{1}{4}v_{\rho }v_{\sigma }\zeta
_{10} & \frac{1}{4}vv_{\sigma }\left( \zeta _{11}-\zeta _{12}\right)%
\end{array}%
\right) , \\
&&D_{3}=\left( 
\begin{array}{ccc}
\frac{1}{4}\left( \beta _{14}v^{2}+2\mu _{\text{r4}}^{2}+v_{\sigma
}^{2}\gamma _{12}\right) & -\frac{A_{7}v_{\sigma }}{2\sqrt{2}} & 0 \\ 
-\frac{A_{7}v_{\sigma }}{2\sqrt{2}} & \frac{1}{4}\left( \kappa _{\text{p1}%
}v^{2}+2\mu _{e}^{2}+v_{\sigma }^{2}\alpha _{\text{e1}}+v_{\rho }^{2}\beta _{%
\text{e3}}\right) & 0 \\ 
0 & 0 & \frac{1}{4}\left( \kappa _{1}v^{2}+\kappa _{2}v^{2}+2\mu _{\text{p2}%
}^{2}+v_{\sigma }^{2}\alpha _{21}+v_{\rho }^{2}\beta _{23}\right)%
\end{array}%
\right) ,  \notag
\end{eqnarray}%
where $M_{\mathrm{CPodd}}^{\left( 1\right) }$ and $M_{\mathrm{CPodd}%
}^{\left( 2\right) }$ are the squared mass matrices for the CP-odd scalars,
neutral and charged under $Z_{4}$, respectively. Note that the squared mass
matrix $M_{\mathrm{CPodd}}^{\left( 1\right) }$ (which is written in the
basis $\left( \func{Im}\left( \phi _{1}^{0}\right) ,\func{Im}\left( \sigma
_{1}\right) \right) $) is exactly zero, since $\func{Im}\left( \phi
_{1}^{0}\right) $ and $\func{Im}\left( \sigma _{1}\right) $ are the
Goldstone bosons associated with the longitudinal components of the $Z$ and $%
Z^{\prime }$ gauge bosons, respectively.

Finally, the squared mass matrix for the charged scalar fields in the basis $%
\left( \phi _{1}^{+},\phi _{2}^{+},\varphi _{3}^{+},\varphi _{1}^{+},\varphi
_{2}^{+},\varphi _{4}^{+},\varphi _{5}^{+}\right) $ reads 
\begin{eqnarray}
M_{C} &=&\left( 
\begin{array}{ccc}
0 & 0_{1\times 2} & 0_{1\times 4} \\ 
0_{2\times 1} & M_{C}^{\left( 1\right) } & 0_{2\times 4} \\ 
0_{4\times 1} & 0_{4\times 2} & M_{C}^{\left( 2\right) }%
\end{array}%
\right) ,\hspace{0.7cm}\hspace{0.7cm}M_{C}^{\left( 2\right) }=\left( 
\begin{array}{cc}
M_{C}^{\left( 2a\right) } & M_{C}^{\left( 2b\right) } \\ 
\left( M_{C}^{\left( 2b\right) }\right) ^{T} & M_{C}^{\left( 2c\right) }%
\end{array}%
\right) ,  \notag \\
M_{C}^{\left( 1\right) } &=&\left( 
\begin{array}{cc}
\frac{1}{2}\left( \kappa _{1}v^{2}+2\kappa _{3}v^{2}+2\mu
_{p2}^{2}+v_{\sigma }^{2}\alpha _{21}+v_{\rho }^{2}\beta _{23}\right) & 
\frac{vA_{2}}{\sqrt{2}} \\ 
\frac{vA_{2}}{\sqrt{2}} & \frac{1}{2}\left( \lambda _{31}v^{2}+2\mu
_{f3}^{2}+\varrho _{33}v_{\rho }^{2}+\varsigma _{31}v_{\sigma }^{2}\right)%
\end{array}%
\right) \,,\hspace{0.7cm}  \notag \\
M_{C}^{\left( 2a\right) } &=&\left( 
\begin{array}{cc}
\frac{1}{2}\left( \lambda _{11}v^{2}+2\mu _{f1}^{2}+\varrho _{13}v_{\rho
}^{2}+\varsigma _{11}v_{\sigma }^{2}\right) & 0 \\ 
0 & \frac{1}{2}\left( \lambda _{21}v^{2}+2\mu _{f2}^{2}+\varrho _{23}v_{\rho
}^{2}+\varsigma _{21}v_{\sigma }^{2}\right)%
\end{array}%
\right) , \\
M_{C}^{\left( 2b\right) } &=&\left( 
\begin{array}{cc}
0 & \frac{\zeta _{4}v_{\sigma }^{2}}{2} \\ 
\frac{A_{9}v_{\rho }}{\sqrt{2}} & \frac{\zeta _{13}v_{\sigma }v_{\rho }}{2}%
\end{array}%
\right) ,\hspace{0.7cm}M_{C}^{\left( 2c\right) }=\left( 
\begin{array}{cc}
\frac{1}{2}\left( \lambda _{41}v^{2}+2\mu _{f4}^{2}+\varrho _{43}v_{\rho
}^{2}+\varsigma _{41}v_{\sigma }^{2}\right) & \frac{A_{3}v_{\sigma }}{\sqrt{2%
}} \\ 
\frac{A_{3}v_{\sigma }}{\sqrt{2}} & \frac{1}{2}\left( \lambda
_{51}v^{2}+2\mu _{f5}^{2}+\varrho _{53}v_{\rho }^{2}+\varsigma
_{51}v_{\sigma }^{2}\right)%
\end{array}%
\right) ,  \notag
\end{eqnarray}%
such that $\phi _{1}^{\pm }$ are the electrically charged massless scalar
states corresponding to the Goldstone bosons associated with the
longitudinal components of the $W^{\pm }$ gauge bosons. 
\begin{figure}[h]
\resizebox{18cm}{22cm}{\includegraphics{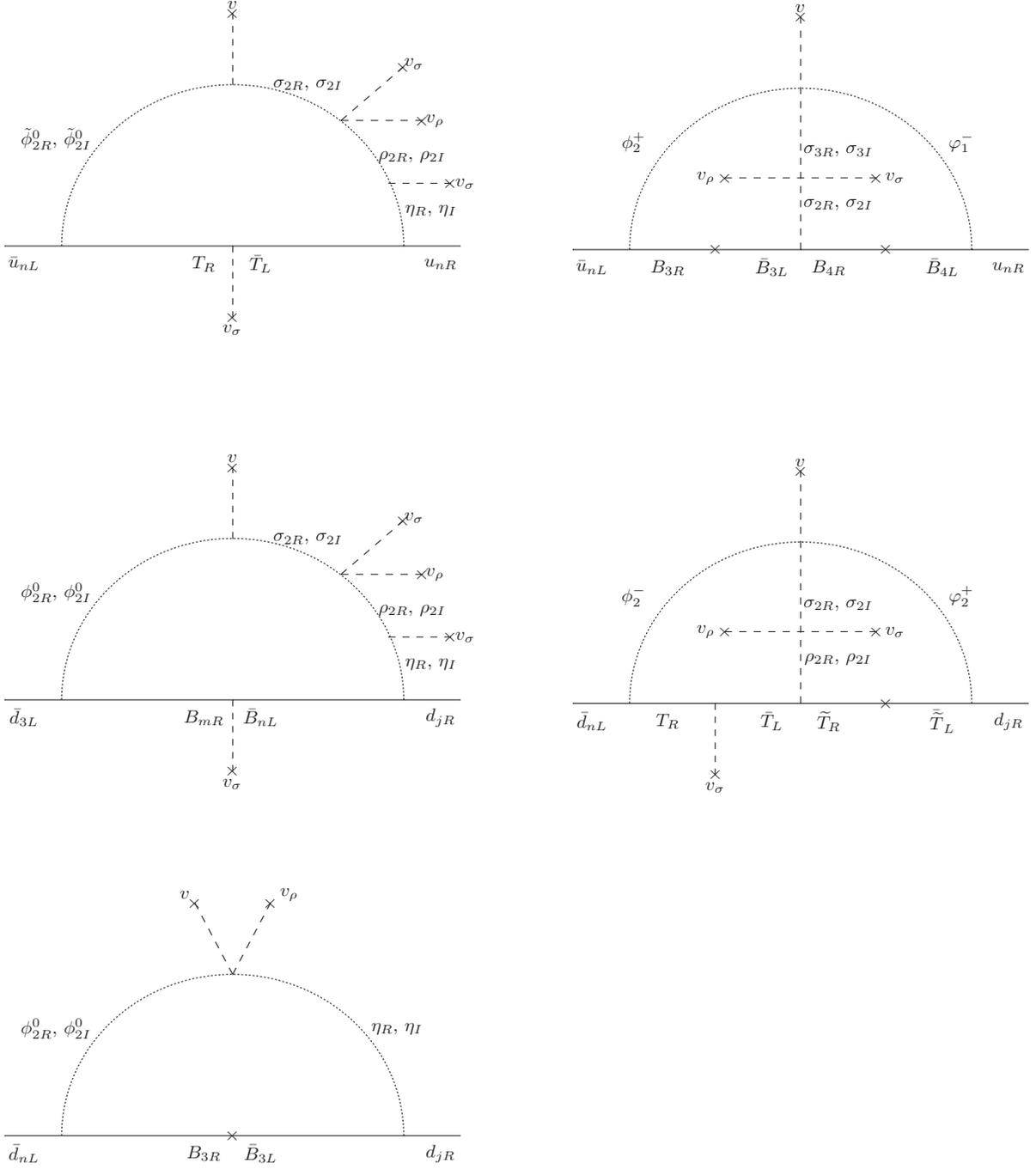}} \vspace{%
-1cm}
\caption{One- and two-loop Feynman diagrams contributing to the entries of
the SM quark mass matrices. Here, $n,m=1,2$ and $j=1,2,3$.}
\label{Loopdiagramsq}
\end{figure}


\section{Radiatively generated quark mass and mixing hierarchies}

\label{Sec:quark-mass-mixing} 

The quark Yukawa interactions determined by Eq.~(\ref{LY}) give rise to the
following up and down mass matrices for the SM quarks, respectively, 
\begin{equation}
M_{U}=\left( 
\begin{array}{ccc}
\varepsilon_{11}^{\left( u\right) }+\widetilde{\varepsilon }%
_{11}^{\left(u\right)} & \varepsilon_{12}^{\left( u\right) }+\widetilde{%
\varepsilon }_{12}^{\left(u\right)} & 0 \\ 
\varepsilon_{21}^{\left( u\right) }+\widetilde{\varepsilon }%
_{21}^{\left(u\right)} & \varepsilon_{22}^{\left( u\right) }+\widetilde{%
\varepsilon }_{22}^{\left(u\right)} & 0 \\ 
0 & 0 & y_{33}^{\left( u\right) }%
\end{array}
\right) \allowbreak \allowbreak \frac{v}{\sqrt{2}},\hspace{1cm}\hspace{1cm}
M_{D}=\left( 
\begin{array}{ccc}
\varepsilon_{11}^{\left( d\right) }+\widetilde{\varepsilon }%
_{11}^{\left(d\right)} & \varepsilon_{12}^{\left( d\right) }+\widetilde{%
\varepsilon }_{12}^{\left(d\right)} & \varepsilon_{13}^{\left( d\right) }+%
\widetilde{\varepsilon }_{13}^{\left(d\right)} \\ 
\varepsilon_{21}^{\left( d\right) }+\widetilde{\varepsilon }%
_{21}^{\left(d\right)} & \varepsilon_{22}^{\left( d\right) }+\widetilde{%
\varepsilon }_{22}^{\left(d\right)} & \varepsilon_{23}^{\left( d\right) }+%
\widetilde{\varepsilon }_{23}^{\left(d\right)} \\ 
\varepsilon_{31}^{\left( d\right) } & \varepsilon_{32}^{\left( d\right) } & 
\varepsilon_{33}^{\left( d\right) }%
\end{array}
\right) \allowbreak \allowbreak \frac{v}{\sqrt{2}} \,,
\end{equation}
where the dimensionless parameters $\varepsilon_{nm}^{\left( u\right) }$ ($%
n,m=1,2$) and $\varepsilon_{ij}^{\left( d\right) }$ ($i,j=1,2,3$) are
generated at one-loop level, whereas $\widetilde{\varepsilon }_{nj}^{\left(
d\right) }$ and $\widetilde{\varepsilon }_{ij}^{\left( d\right) }$ arise at
two-loop level. The characteristic Feynman loop diagrams contributing to the
entries of the SM quark mass matrices are shown in Fig.~\ref{Loopdiagramsq}.

In what follows, we demonstrate that the mass matrices for SM quarks given
above incorporate the observed hierarchies in the SM quark mass spectrum and
the Cabibbo-Kobayashi-Maskawa (CKM) mixing matrix. To this end, we proceed
with a parametrization of the SM quark mass matrices in the following form 
\begin{equation}
M_{U}=\left( 
\begin{array}{ccc}
\left( a_{11}^{\left( u\right) }\right) ^{2}l & \left( a_{12}^{\left(
u\right) }\right) ^{2}l & 0 \\ 
\left( a_{21}^{\left( u\right) }\right) ^{2}l & \left( a_{22}^{\left(
u\right) }\right) ^{2}l & 0 \\ 
0 & 0 & y_{33}^{\left( u\right) }%
\end{array}%
\right) \allowbreak \allowbreak \frac{v}{\sqrt{2}},\hspace{1cm}\hspace{1cm}%
M_{D}=\left( 
\begin{array}{ccc}
\left( a_{11}^{\left( d\right) }\right) ^{2}l & \left( a_{12}^{\left(
d\right) }\right) ^{2}l & \left( a_{13}^{\left( d\right) }\right) ^{2}l \\ 
\left( a_{21}^{\left( d\right) }\right) ^{2}l & \left( a_{22}^{\left(
d\right) }\right) ^{2}l & \left( a_{23}^{\left( d\right) }\right) ^{2}l \\ 
\left( a_{31}^{\left( d\right) }\right) ^{2}l & \left( a_{32}^{\left(
d\right) }\right) ^{2}l & \left( a_{33}^{\left( d\right) }\right) ^{2}l%
\end{array}%
\right) \allowbreak \allowbreak \frac{v}{\sqrt{2}}\,,
\end{equation}%
where $l\approx (1/4\pi )^{2}\approx 2.0\times \lambda ^{4}$ is the loop
suppression factor, and $\lambda =0.225$ is the Wolfenstein parameter. As a
consequence, we expect that $a_{nm}^{\left( u\right) }$, $a_{ij}^{\left(
d\right) }$ ($n,m=1,2$ and $i,j=1,2,3$) be $\mathcal{O}(1)$ parameters.

We remark that the Feynman diagrams contributing to the entries of the SM
fermion mass matrices contain a large number of uncorrelated parameters that
belong to the fermion and scalar sectors of our model. Nevertheless, these
parameters can be absorbed into a limited number of effective parameters $%
\varepsilon_{nm}^{\left( u\right) }$, $\widetilde{\varepsilon }_{nj}^{\left(
d\right) }$, $\varepsilon_{ij}^{\left( d\right) }$, $\widetilde{\varepsilon }%
_{ij}^{\left( d\right) }$ ($n,m=1,2$ and $i,j=1,2,3$), which can be used to
reproduce the experimental values of the physical observables in the quark
sector\footnote{%
We use the experimental values of the quark masses at the $M_{Z}$ scale
known from Ref.~\cite{Bora:2012tx}, which are similar to those in Ref.~\cite%
{Xing:2007fb}. The experimental values of the CKM parameters are taken from
Ref.~\cite{Olive:2016xmw}.} 
\begin{eqnarray}
&& m_{u}(MeV)=1.45_{-0.45}^{+0.56}, \hspace{3mm}
m_{d}(MeV)=2.9_{-0.4}^{+0.5}, \hspace{3mm} m_{s}(MeV)=57.7_{-15.7}^{+16.8},
\label{eq:Qsector-observables} \\
&&m_{c}(MeV)=635\pm 86,\hspace{3mm} m_{t}(GeV)=172.1\pm 0.6\pm 0.9,\hspace{%
3mm} m_{b}(GeV)=2.82_{-0.04}^{+0.09},\hspace{3mm}  \notag \\
&&\sin \theta_{12}=0.2254,\hspace{3mm} \sin \theta_{23}=0.0414,\hspace{3mm}
\sin \theta_{13}=0.00355,\hspace{3mm} J=2.96_{-0.16}^{+0.20}\times 10^{-5}
\,.  \notag
\end{eqnarray}%
Here, $m_{t,u,c,d,s,b}$ are the SM quark masses, $\theta_{12}$, $\theta_{23}$%
, $\theta_{13}$ are the mixing angles, and $J$ is the Jarlskog parameter.

While our model does not predict the exact values of the physical
observables, it offers a natural explanation of the observed (strong)
hierarchies. As was previously mentioned, it only pretends to reproduce the
existing pattern of quark masses and mixing caused by a sequential loop
suppression predicted by the model. To this end, for the SM quark mass
matrices given above, we look for the eigenvalue problem solutions
reproducing the experimental values of the quark masses and the CKM
parameters given by Eq.~(\ref{eq:Qsector-observables}), requiring that $%
a^{(u,d)},b^{(u,d)}$ are all of the same order of one. The standard
procedure renders the following solution 
\begin{eqnarray}
a_{11}^{\left( u\right) } &\simeq &0.708,\hspace{1cm} a_{12}^{\left(
u\right) } = a_{21}^{\left( u\right) }\simeq 0.567,\hspace{1cm}
a_{22}^{\left( u\right) } \simeq 0.456,\hspace{1cm}y_{33}^{\left( u\right)
}=0.989,  \notag \\
a_{11}^{\left( d\right) } &\simeq& 0.191,\hspace{1cm} a_{12}^{\left(
d\right)} = a_{21}^{\left( d\right) }\simeq 0.182,\hspace{1cm}
a_{13}^{\left( d\right)} = a_{31}^{\left( d\right) }\simeq 0.325+0.009i, 
\notag \\
a_{23}^{\left( d\right) } &=& a_{32}^{\left( d\right) }\simeq 0.269-0.016i,%
\hspace{1cm} a_{22}^{\left( d\right) }\simeq 0.190,\hspace{1cm}
a_{33}^{\left(d\right) }\simeq 1.771 \,.
\end{eqnarray}%
The above $\mathcal{O}(1)$ values exactly reproduce the measured central
values of the SM quark masses and CKM parameters given in Eq.~(\ref%
{eq:Qsector-observables}). Hence, our model is consistent with and
successfully reproduces the existing pattern of SM quark masses caused by
the sequential loop suppression mechanism, with different quark flavors
getting mass at different orders in Perturbation Theory as discussed above. 
\begin{figure}[!h]
\resizebox{18cm}{22cm}{\includegraphics{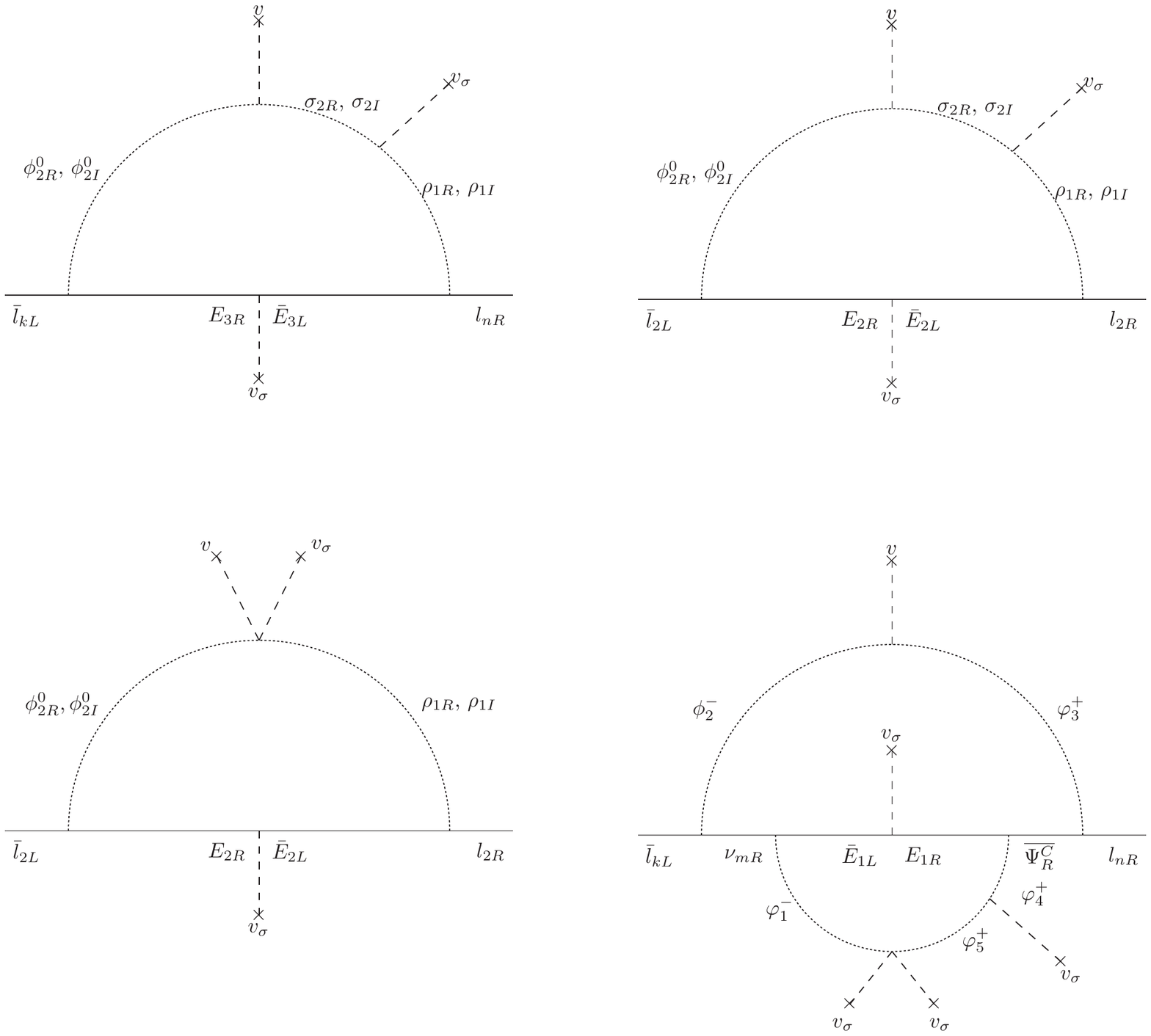}} 
\vspace{-7.5cm}
\caption{One- and two-loop Feynman diagrams contributing to the entries of
the SM charged lepton mass matrix. Here, $k,m,n=1,3$.}
\label{Loopdiagramsl}
\end{figure}

\begin{figure}
 \centering
    \subfigure[]
{\resizebox{18cm}{22cm}{\includegraphics{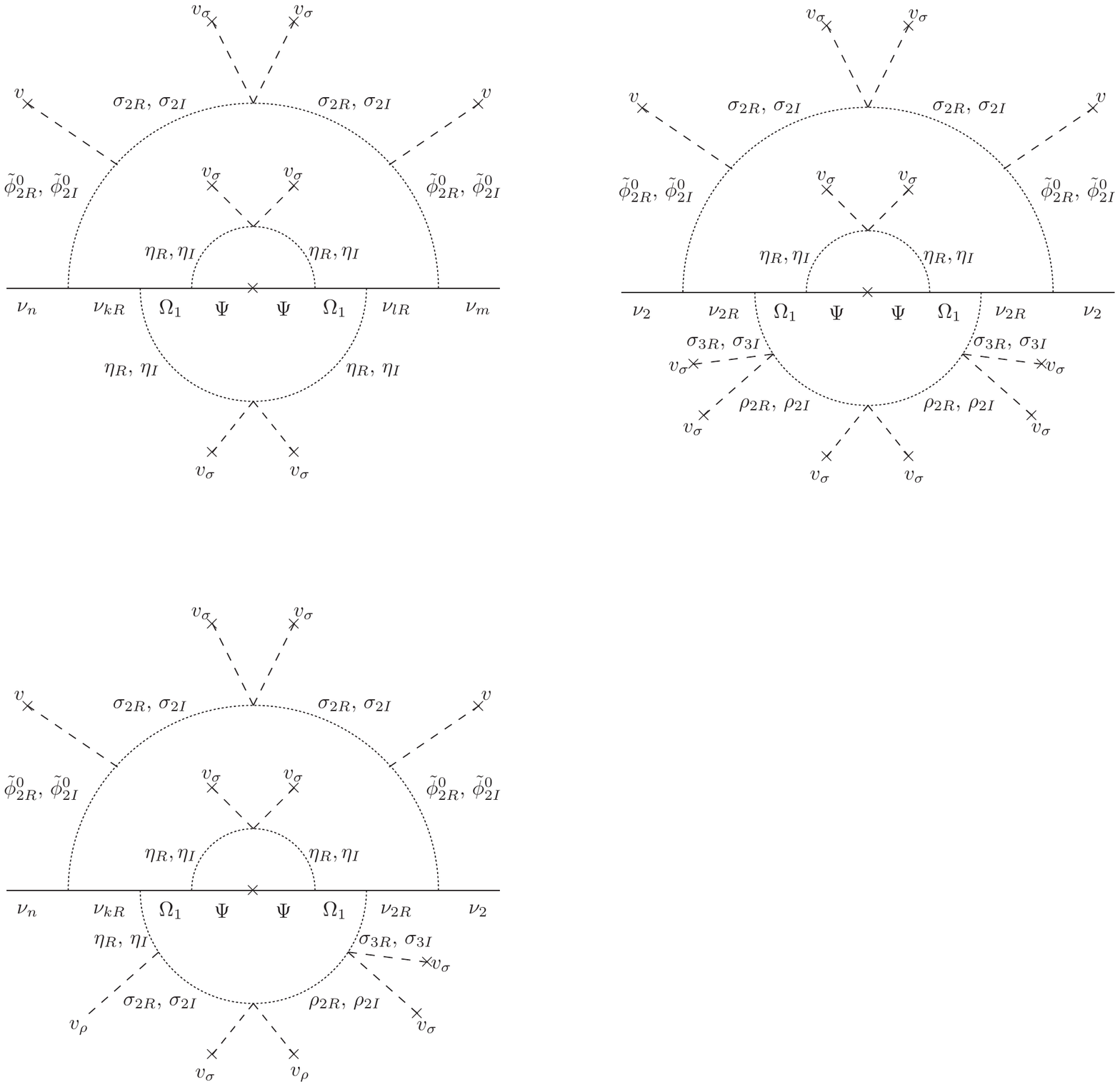}}}
 \vspace{-15cm}
    \subfigure[]
{\hspace{8cm}\resizebox{11cm}{14cm}{\includegraphics{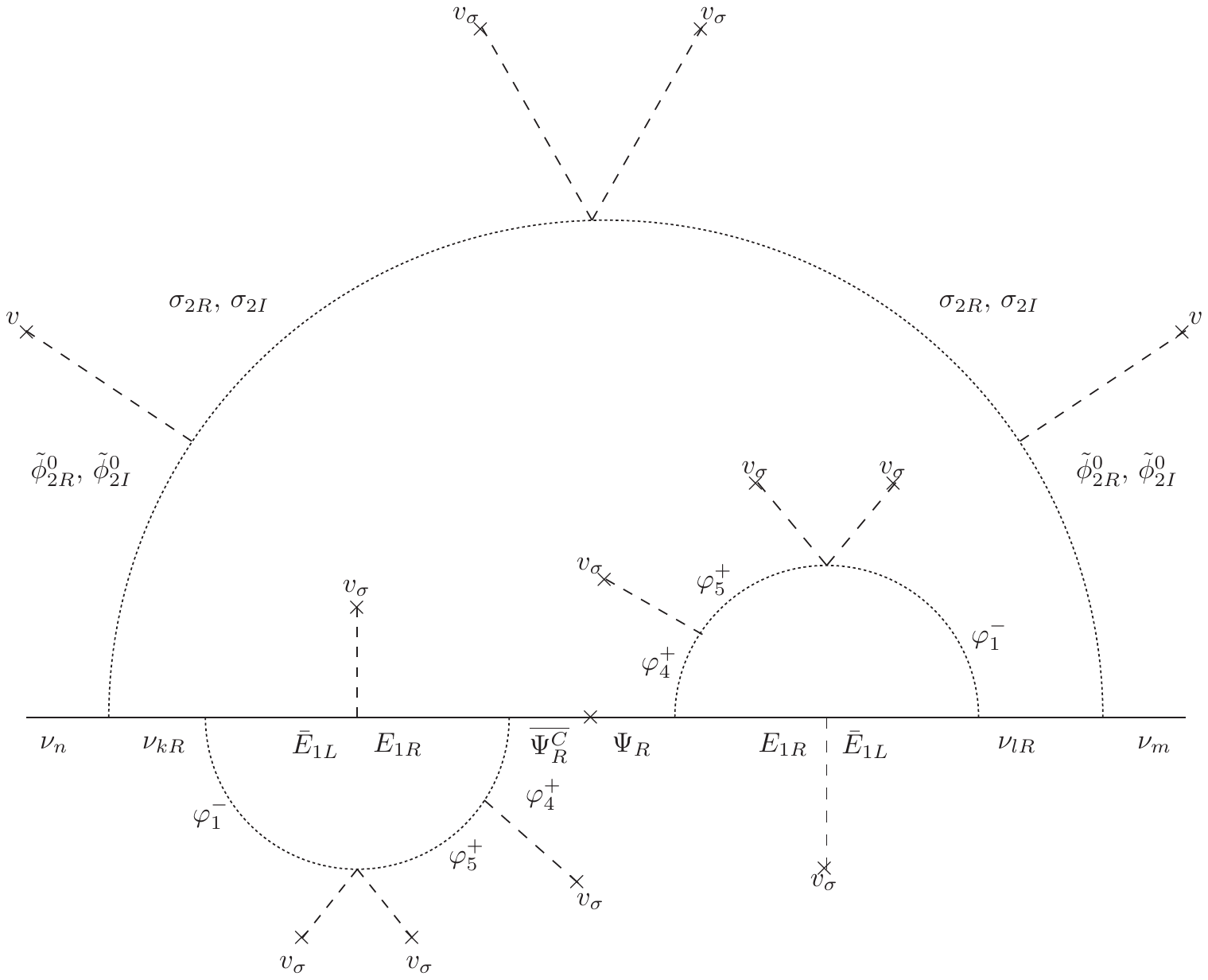}}}
 \vspace{-6cm}
\caption{Three-loop Feynman diagrams contributing to the entries of the
neutrino mass matrix. Here, $k,l,n,m=1,3$.}\vspace{12cm}
\label{Loopdiagramsnu}
\end{figure}


\section{Radiatively generated lepton masses and mixings}

\label{Sec:lepton-mass-mixing} 

The lepton and neutrino Yukawa interactions and mass terms given in Eq.~(\ref%
{LY}) give rise to the characteristic Feynman loop diagrams illustrated in
Figs.~\ref{Loopdiagramsl} and \ref{Loopdiagramsnu} that necessarily generate
the following SM charged lepton and light active neutrino mass forms: 
\begin{equation}
M_{l}=\left( 
\begin{array}{ccc}
\varepsilon _{11}^{\left( l\right) }+\widetilde{\varepsilon }_{11}^{\left(
l\right) } & 0 & \varepsilon _{13}^{\left( l\right) }+\widetilde{\varepsilon 
}_{13}^{\left( l\right) } \\ 
0 & \varepsilon _{22}^{\left( l\right) } & 0 \\ 
\varepsilon _{31}^{\left( l\right) }+\widetilde{\varepsilon }_{31}^{\left(
l\right) } & 0 & \varepsilon _{33}^{\left( l\right) }+\widetilde{\varepsilon 
}_{33}^{\left( l\right) }%
\end{array}%
\right) \allowbreak \allowbreak \frac{v}{\sqrt{2}},\hspace{1cm}\hspace{1cm}%
M_{\nu }=\left( 
\begin{array}{ccc}
a_{11}^{\left( \nu \right) } & a_{12}^{\left( \nu \right) } & a_{13}^{\left(
\nu \right) } \\ 
a_{21}^{\left( \nu \right) } & a_{22}^{\left( \nu \right) } & a_{23}^{\left(
\nu \right) } \\ 
a_{31}^{\left( \nu \right) } & a_{32}^{\left( \nu \right) } & a_{33}^{\left(
\nu \right) }%
\end{array}%
\right) \allowbreak \allowbreak ,
\end{equation}%
where $\varepsilon _{ii}^{\left( l\right) }$, $\varepsilon _{13}^{\left(
l\right) }$, $\varepsilon _{31}^{\left( l\right) }$ are the dimensionless
parameters generated at one-loop level, whereas the parameters $\widetilde{%
\varepsilon }_{ii}^{\left( d\right) }$, $\widetilde{\varepsilon }%
_{13}^{\left( d\right) }$ and $\widetilde{\varepsilon }_{31}^{\left(
d\right) }$ appear at two-loop level. In what follows, we show that the
lepton mass matrices given above can accommodate the experimental data on
the SM lepton masses and mixing. For this purpose, following the same
strategy as for the quark mass forms discussed in the previous section, we
parametrize the SM charged lepton mass matrix as follows: 
\begin{equation}
M_{l}=\left( 
\begin{array}{ccc}
\left( a_{11}^{\left( l\right) }\right) ^{2}l & 0 & \left( a_{13}^{\left(
l\right) }\right) ^{2}l \\ 
0 & \left( a_{22}^{\left( l\right) }\right) ^{2}l & 0 \\ 
\left( a_{31}^{\left( l\right) }\right) ^{2}l & 0 & \left( a_{33}^{\left(
l\right) }\right) ^{2}l%
\end{array}%
\right) \allowbreak \allowbreak \frac{v}{\sqrt{2}}\,.
\end{equation}%
In order to fit the measured values of the charged lepton masses, as well as
the neutrino mass squared differences and lepton mixing parameters \cite%
{deSalas:2017kay}, we proceed by solving the eigenvalue problem for the SM
lepton and light neutrino mass matrices. The following solution has been
found: 
\begin{eqnarray}
a_{11}^{\left( l\right) } &=&a_{33}^{\left( l\right) }\simeq 0.491,\hspace{%
1cm}a_{13}^{\left( l\right) }=a_{31}^{\left( l\right) }\simeq 0.4905,\hspace{%
1cm}a_{22}^{\left( l\right) }\simeq 0.340, \\
M_{\nu } &=&\left\{ 
\begin{array}{l}
\left( 
\begin{array}{ccc}
0.0473664\,-0.00675494i & 0.00904226\,-0.0045673i & 0.00544117\,+0.000841261i
\\ 
0.00904226\,-0.0045673i & 0.0528446\,+0.000500202i & 0.0066006\,+0.00515561i
\\ 
0.00544117\,+0.000841261i & 0.0066006\,+0.00515561i & 0.04306\,+0.00575822i
\\ 
&  & 
\end{array}%
\right) \mbox{eV}\ \ \ \ \ \mbox{for NH}, \\ 
\\ 
\left( 
\begin{array}{ccc}
0.0573349\,-0.0100651i & -0.00894313-0.00802622i & -0.00466139-0.000301301i
\\ 
-0.00894313-0.00802622i & 0.0495729\,-0.000243999i & -0.00936421+0.00765498i
\\ 
-0.00466139-0.000301301i & -0.00936421+0.00765498i & 0.0560537\,+0.0101163i
\\ 
&  & 
\end{array}%
\right) \mbox{eV}\ \ \ \mbox{for IH}.%
\end{array}%
\right.
\end{eqnarray}%
Thus we find that with $\mathcal{O}(1)$ values for the $a_{nm}^{\left(
l\right) }$ ($n,m=1,3$), $a_{22}^{\left( l\right) }$ parameters and the
above specified entries of the neutrino mass matrix satisfying $\mathcal{O}%
(10^{-3})$ eV$\lesssim \left\vert \left( M_{\nu }\right) _{ij}\right\vert
\lesssim \mathcal{O}(10^{-2})$ eV ($i,j=1,2,3$), the experimental values for
the physical observables of the lepton sector, i.e., the three charged
lepton masses, the two neutrino mass squared splittings, the three leptonic
mixing parameters and the leptonic Dirac CP violating phase can be
successfully reproduced. Consequently, our model is consistent with and
successfully reproduces the existing pattern of SM charged lepton masses
generated by the sequential loop suppression mechanism.


\section{Discussions and Conclusions}

\label{Sec:conclusions} 

We have constructed a first renormalizable extension of the Inert Doublet
Model that enables an implementation of a sequential loop suppression
mechanism, capable of explaining the observed SM fermion mass hierarchy
without invoking soft breaking mass terms. In our model, the SM gauge
symmetry is supplemented by the $U_{1X}\times Z_{2}^{\left( 1\right) }\times
Z_{2}^{\left( 2\right) }$ family symmetry, where the gauge $U_{1X}$ and
discrete $Z_{2}^{\left( 1\right) }$ symmetries are spontaneously broken,
whereas the $Z_{2}^{\left( 2\right) }$ symmetry is preserved. 

Our model is consistent with the observed SM fermion mass spectrum and fermionic mixing
parameters and allows for an explanation of the recently observed $R_{K}$
and $R_{K^{\ast }}$ anomalies, thanks to the non-universal $Z^{\prime }$
couplings to fermions.
Let us point out that in the case of the studied sequential loop mechanism these anomalies cannot be explained by the Yukawa couplings.
Indeed, as follows from Eq. (\ref{LY}), there are no terms of the form $\overline{f}_{iL}Sf_{jR}$ (i,j=1,2,3), where  $S$ is a scalar and $f_i$ denote the SM fermions. Consequently, at tree level there is no scalar exchange contribution to the $R_{K,K^{\ast}}$ anomalies. 
This contribution appears only at two loop level given that interactions of the form $\overline{b}_{L}\sigma\overline{s}_{R}$, $\overline{e}_{L}\sigma\overline{e}_{R}$ and $\overline{\mu}_{L}\sigma\overline{\mu}_{R}$ are generated at one loop level as seen from 
the one loop  diagrams for the down type quark and the charged leptons masses shown in Figs.~\ref{Loopdiagramsq} and \ref{Loopdiagramsl}. Thus, the Yukawa contributions to the $R_{K,K^{\ast}}$ anomalies are very much suppressed by the loop factors and the heavy particles in the loops.
%

We focused on an extension of the Inert Higgs Doublet model (IDM) that
allows the implementation of the sequential loop suppression mechanism for
the generation of SM fermion masses instead of an extension of the inert
3-3-1 model (model based on the $SU_{3C}\times SU_{3L}\times U_{1X}$ gauge
symmetry). As previously mentioned, the extension of the inert 3-3-1 model
of Ref~\cite{CarcamoHernandez:2017cwi} does not explain the $R_{K}$ and $%
R_{K^{\ast }}$ anomalies and the light active neutrino masses appear at
two-loop level like the masses of the light SM charged fermions. Addressing
the $R_{K}$ and $R_{K^{\ast }}$ anomalies in the framework of a 3-3-1 model
would require to consider five families of $SU(3)_L$ leptonic triplets as
done in Ref. \cite{Descotes-Genon:2017ptp}, in order to have different $%
U(1)_X$ charge assignments for the first and second lepton families, without
spoiling the anomaly cancellation conditions. Thus, modifying the inert
3-3-1 model of Ref~\cite{CarcamoHernandez:2017cwi} to account for the $R_{K}$
and $R_{K^{\ast }}$ anomalies, and to generate the hierarchy of SM fermion
masses by sequential loop suppression mechanism, with the light active
neutrino masses appearing at three-loop level, will require a much larger
particle content than the one adopted in the framework of an extended IDM.

In our model only the top quark and exotic fermions acquire tree-level
masses, whereas the masses of the remaining SM fermions emerge from a
radiative seesaw-like mechanism: the masses for the bottom, strange and
charm quarks, tau and muon leptons are generated at one-loop level, whereas
the masses for the up and down quarks as well as the electron mass appear at
two-loop level. Furthermore, light active neutrino acquire masses by means
of a radiative seesaw mechanism at three-loop level.

Due to an unbroken $Z_{2}^{\left( 2\right) }$ discrete symmetry, our model
has several stable scalar dark matter candidates, which can be the neutral
components of the inert $SU_{2L}$ scalar doublet $\phi _{2}$ as well as the
real and imaginary parts of the SM scalar singlets $\sigma _{2}$, $\sigma
_{3}$, $\rho _{1}$, $\rho _{2}$ and $\eta $. Furthermore, the model can have
a fermionic dark matter candidate which is the only SM-singlet Majorana
neutrino $\Omega _{1R}$ with a non-trivial $Z_{2}^{\left( 2\right) }$
charge. A study of the phenomenological implications of our model goes
beyond the scope of the present paper and will be performed in a forthcoming
work.

\textbf{Acknowledgements} 
This research has received funding from Fondecyt (Chile) grants No.~1170803,
No.~1190845, No.~1180232, No.~3150472 and by CONICYT (Chile) PIA/Basal
FB0821, the UTFSM grant FI4061. R.P. is partially supported by the Swedish
Research Council, contract numbers 621-2013-4287 and 2016-05996, by CONICYT
grant MEC80170112, as well as by the European Research Council (ERC) under
the European Union's Horizon 2020 research and innovation programme (grant
agreement No 668679). This work was supported in part by the Ministry of
Education, Youth and Sports of the Czech Republic, project LTC17018. A.E.C.H
thanks University of Lund, where part of this work was done, for hospitality
and University of Southampton for hospitality during the completion of this
work.

\end{document}